\documentclass[fleqn,usenatbib]{mnras}

\usepackage{newtxtext,newtxmath}

\usepackage[T1]{fontenc}

\DeclareRobustCommand{\VAN}[3]{#2}
\let\VANthebibliography\thebibliography
\def\thebibliography{\DeclareRobustCommand{\VAN}[3]{##3}\VANthebibliography}

\usepackage{graphicx}	
\usepackage{amsmath}	
\usepackage{xcolor}
\usepackage{ulem}

\newcommand\boryana[1]{\textcolor{black}{#1}}
\def\mpcoh{\,h^{-1}{\rm Mpc}} 
\def\gpcoh{\,h^{-1}{\rm Gpc}} 
\def\hompc{\,h\,{\rm Mpc}^{-1}}
\def\msunoh{\,h^{-1}{\rm M}_\odot}

\title[\textsc{AbacusSummit} weak lensing catalogues]{Synthetic light cone catalogues of modern redshift and weak lensing surveys with \textsc{AbacusSummit}}

\author[B. Hadzhiyska et al.]{Boryana~Hadzhiyska\thanks{E-mail: boryanah@berkeley.edu},$^{1,2,3}$ 
S.~Yuan,$^{3}$
C.~Blake,$^{4}$
D.~J.~Eisenstein,$^{5}$
J.~Aguilar,$^{1}$
S.~Ahlen,$^{6}$
D.~Brooks,$^{7}$
\newauthor
T.~Claybaugh,$^{1}$
A.~de la Macorra,$^{8}$
P.~Doel,$^{7}$
N.~Emas,$^{4}$
J.~E.~Forero-Romero,$^{9}$
C.~Garcia-Quintero,$^{10}$
M.~Ishak,$^{10}$
\newauthor
S.~Joudaki,$^{11}$
E.~Jullo,$^{12}$
R.~Kehoe,$^{13}$
T.~Kisner,$^{1}$
A.~Kremin,$^{1}$
A.~Krolewski,$^{11,14,15}$
M.~Landriau,$^{1}$
J.~U.~Lange,$^{16,17}$
\newauthor
M.~Manera,$^{18}$
R.~Miquel,$^{19,18}$
Jundan Nie,$^{20}$
C.~Poppett,$^{1,21,2}$
A.~Porredon,$^{22,23}$
G.~Rossi,$^{24}$
R.~Ruggeri,$^{4,25}$
\newauthor
C.~Saulder,$^{26}$
M.~Schubnell,$^{16,17}$
G.~Tarl\'{e},$^{17}$
B.~A.~Weaver,$^{27}$
E.~Xhakaj,$^{28}$
and Zhimin~Zhou$^{20}$
\\
$^{1}$ Lawrence Berkeley National Laboratory, 1 Cyclotron Road, Berkeley, CA 94720, USA\\
$^{2}$ University of California, Berkeley, 110 Sproul Hall \#5800 Berkeley, CA 94720, USA\\
$^{3}$ SLAC National Accelerator Laboratory, Menlo Park, CA 94305, USA\\
$^{4}$ Centre for Astrophysics \& Supercomputing, Swinburne University of Technology, P.O. Box 218, Hawthorn, VIC 3122, Australia\\
$^{5}$ Center for Astrophysics $|$ Harvard \& Smithsonian, 60 Garden Street, Cambridge, MA 02138, USA\\
$^{6}$ Physics Dept., Boston University, 590 Commonwealth Avenue, Boston, MA 02215, USA\\
$^{7}$ Department of Physics \& Astronomy, University College London, Gower Street, London, WC1E 6BT, UK\\
$^{8}$ Instituto de F\'{\i}sica, Universidad Nacional Aut\'{o}noma de M\'{e}xico,  Cd. de M\'{e}xico  C.P. 04510,  M\'{e}xico\\
$^{9}$ Departamento de F\'isica, Universidad de los Andes, Cra. 1 No. 18A-10, Edificio Ip, CP 111711, Bogot\'a, Colombia\\
$^{10}$ Department of Physics, The University of Texas at Dallas, Richardson, TX 75080, USA\\
$^{11}$ Department of Physics and Astronomy, University of Waterloo, 200 University Ave W, Waterloo, ON N2L 3G1, Canada\\
$^{12}$ Aix Marseille Univ, CNRS, CNES, LAM, Marseille, France\\
$^{13}$ Department of Physics, Southern Methodist University, 3215 Daniel Avenue, Dallas, TX 75275, USA\\
$^{14}$ Perimeter Institute for Theoretical Physics, 31 Caroline St. North, Waterloo, ON N2L 2Y5, Canada\\
$^{15}$ Waterloo Centre for Astrophysics, University of Waterloo, 200 University Ave W, Waterloo, ON N2L 3G1, Canada\\
$^{16}$ Department of Physics, University of Michigan, Ann Arbor, MI 48109, USA\\
$^{17}$ University of Michigan, Ann Arbor, MI 48109, USA\\
$^{18}$ Institut de F\'{i}sica d’Altes Energies (IFAE), The Barcelona Institute of Science and Technology, Campus UAB, 08193 Bellaterra Barcelona, Spain\\
$^{19}$ Instituci\'{o} Catalana de Recerca i Estudis Avan\c{c}ats, Passeig de Llu\'{\i}s Companys, 23, 08010 Barcelona, Spain\\
$^{20}$ National Astronomical Observatories, Chinese Academy of Sciences, A20 Datun Rd., Chaoyang District, Beijing, 100012, P.R. China\\
$^{21}$ Space Sciences Laboratory, University of California, Berkeley, 7 Gauss Way, Berkeley, CA  94720, USA\\
$^{22}$ Institute for Astronomy, University of Edinburgh, Royal Observatory, Blackford Hill, Edinburgh EH9 3HJ, UK\\
$^{23}$ The Ohio State University, Columbus, 43210 OH, USA\\
$^{24}$ Department of Physics and Astronomy, Sejong University, Seoul, 143-747, Korea\\
$^{25}$ School of Mathematics and Physics, University of Queensland, 4072, Australia\\
$^{26}$ Korea Astronomy and Space Science Institute, 776, Daedeokdae-ro, Yuseong-gu, Daejeon 34055, Republic of Korea\\
$^{27}$ NSF's NOIRLab, 950 N. Cherry Ave., Tucson, AZ 85719, USA\\
$^{28}$ Department of Astronomy and Astrophysics, University of California, Santa Cruz, 1156 High Street, Santa Cruz, CA 95064 USA\\
}
\date{Accepted XXX. Received YYY; in original form ZZZ}

\pubyear{2015}

\begin{document}
\label{firstpage}
\pagerange{\pageref{firstpage}--\pageref{lastpage}}
\maketitle

\begin{abstract}
The joint analysis of different cosmological probes, such as galaxy clustering and weak lensing, can potentially yield invaluable insights into the nature of the primordial Universe, dark energy and dark matter. However, the development of high-fidelity theoretical models that cover a wide range of scales and redshifts is a necessary stepping-stone. Here, we present public high-resolution weak lensing maps on the light cone, generated using the $N$-body simulation suite \textsc{AbacusSummit} in the Born approximation, and accompanying weak lensing mock catalogues, tuned via fits to the Early Data Release small-scale clustering measurements of the Dark Energy Spectroscopic Instrument (DESI). Available in this release are maps of the cosmic shear, deflection angle and convergence fields at source redshifts ranging from $z = 0.15$ to 2.45 with $\Delta z = 0.05$ as well as CMB convergence maps ($z \approx 1090$) for each of the 25 \texttt{base}-resolution simulations ($L_{\rm box} = 2000\,h^{-1}{\rm Mpc}$, $N_{\rm part} = 6912^3$) as well as for the two \texttt{huge} simulations ($L_{\rm box} = 7500\,h^{-1}{\rm Mpc}$, $N_{\rm part} = 8640^3$) at the fiducial AbacusSummit cosmology ($Planck$ 2018). The pixel resolution of each map is 0.21 arcmin, corresponding to a HEALPiX $N_{\rm side}$ of 16384. The sky coverage of the \texttt{base} simulations is an octant until $z \approx 0.8$ (decreasing to about 1800 deg$^2$ at $z \approx 2.4$), whereas the \texttt{huge} simulations offer full-sky coverage until $z \approx 2.2$. Mock lensing source catalogues are sampled matching the ensemble properties of the Kilo-Degree Survey, Dark Energy Survey, and Hyper-Suprime Cam weak lensing datasets. The produced mock catalogues are validated against theoretical predictions for various clustering and lensing statistics such as galaxy clustering multipoles, galaxy-shear and shear-shear, showing excellent agreement. All products can be downloaded \href{https://app.globus.org/file-manager?origin_id=3dd6566c-eed2-11ed-ba43-09d6a6f08166&path=%2F}{via a Globus endpoint} (see Data Availability).
\end{abstract}

\begin{keywords}
keyword1 -- keyword2 -- keyword3
\end{keywords}



\section{Introduction}
\label{sec:intro}
The quest to understand dark energy and pin down possible deviations from the standard model has spawned a number of large-scale structure (LSS) experiments. Measuring the statistics of LSS provides a powerful tool for constraining dark energy, which complements Type Ia supernovae and cosmic microwave background (CMB) probes. The galaxy two-point correlation function, which describes the spatial clustering of galaxies, has provided some of the earliest and most robust evidence for the $\Lambda$CDM model \citep{2005MNRAS.362..505C,2006PhRvD..74l3507T,1977ApJ...217..385G,1984Natur.311..517B,1990MNRAS.243..692M,1996MNRAS.280..267B,1996MNRAS.283.1227M,2001ApJ...546....2E,1992MNRAS.254..295C,1998MNRAS.300..493S,2001ApJ...555..547H,2000MNRAS.317...55S,2002MNRAS.330..506H,2001MNRAS.327.1297P,2007MNRAS.378..852P}. In addition, LSS is sensitive to dark matter and dark energy through cosmic shear -- slight distortions of the shapes of background galaxies due to the gravitational lensing of the light rays traveling towards us. A benefit of cosmic shear measurements is that they are more directly related to the distribution of mass and can thus be used to stress test the cosmological paradigm. Finally, the cross-correlation between lens galaxy positions and source galaxy shapes, known as galaxy–galaxy lensing, provides a link between galaxy clustering and cosmic shear and a powerful tool for the joint analysis of weak lensing and galaxy clustering measurements, which is known to break degeneracies between a number of model parameters, thereby harvesting more constraining power for the cosmological parameters \citep{1996ApJ...466..623B,2000AJ....120.1198F,2006MNRAS.368..715M,2007arXiv0709.1159J,2009MNRAS.394..929C,2012ApJ...744..159L,2012ApJ...759..101C,2013MNRAS.432.1544M,2014MNRAS.437.2111V,2017MNRAS.465.4204C,2017MNRAS.467.3024L,2017MNRAS.464.4045K,2018PhRvD..98d2005P}. The combination of these LSS probes: galaxy clustering, cosmic shear, and galaxy–galaxy lensing, informs us about both structure formation in the late Universe and helps us calibrate poorly understood astrophysical processes.

To meet the science goals of the new generation of surveys (DESI, {\it Euclid}, DES, KiDS, HSC, LSST) \citep{2016arXiv161100036D,2019BAAS...51g..57L,2015AJ....150..150F,2018PhRvD..98d3526A,2016MNRAS.460.1270D,2019PASJ...71...43H,2018PASJ...70S...4A,2021A&A...646A.140H,2015MNRAS.454.3500K,2013LRR....16....6A,2012arXiv1211.0310L}, it is crucial to match our experimental efforts with theoretical ones. However, the lensing correlations on sub-degree angular scales are the result of non-linear gravitational clustering, which cannot be described by a purely analytic prescription. In lieu of a completely analytic model, cosmologists often adopt numerical $N$-body simulations, which through clever computational techniques, arrive at the answer for the non-linear growth of large-scale structures, and can thus be used to model the gravitational lensing distortions. Traditionally, producing weak lensing observables via numerical simulations relies on ray-tracing techniques \citep{1991MNRAS.251..600B,1998ApJ...494...29W,2000ApJ...530..547J,2000ApJ...537....1W,2002MNRAS.330..365H,2003ApJ...592..699V,2004APh....22...19W,2009A&A...499...31H,2013MNRAS.435..115B}. In this approach, light rays are back-traced from the observer to the source, as they are deflected from multiple projected-mass lens planes. Statistics measured from ray-tracing simulations have shown good agreement with the predictions of non-linear theory, but they come at a substantial computational price and are thus often restricted to small (few deg$^2$) patches in the sky.

In order to accurately model the wide-field measurements of current experiments, one needs to cover a wide range of scales: from the few-degree large linear scales down to the few-arcmin small non-linear scales.
An alternative method to ray-tracing, which is much less computationally expensive and thus readily capable of producing large lensing maps, involves the so-called Born approximation, where the matter from an $N$-body simulation is projected along unperturbed paths using the single-plane approximation. This method can be implemented over large-volume high-resolution simulations and and is well-known to yield accurate weak-lensing observables on the curved sky for sources at $z_s \sim 3$ or the CMB lensing (for $z_s \approx 1100$) \citep{1998A&A...331..829G,2008MNRAS.391..435F,2008ApJ...682....1D,2009A&A...497..335T}.

Apart from the method of producing weak lensing maps, one needs to also decide on a ``galaxy-painting'' technique for generating mock catalogues with realistic galaxy populations in an $N$-body simulation. Several well-known galaxy population mechanisms are often adopted: the halo occupation distribution (HOD) \cite[e.g.][]{2000MNRAS.318.1144P,2000MNRAS.318..203S,2001ApJ...546...20S,2002ApJ...575..587B,2004ApJ...609...35K,2005ApJ...633..791Z}, which describes the probability a halo with mass $M_{\rm halo}$ contains $N_g$ galaxies; subhalo abundance matching (SHAM) \cite[e.g.][]{2004MNRAS.353..189V,2007ApJ...668..826C}, which relates directly subhalo properties (such as mass and circular velocity) to galaxy properties (such as luminosity and stellar mass); and semi-analytic models (SAMs) \cite[e.g.][]{2006RPPh...69.3101B,2012NewA...17..175B,2008MNRAS.391..481S}, which uses analytic prescriptions to model the formation and evolution of galaxies by usually utilizing the halo merger histories. A prerequisite for both SAMs and SHAM models is the existence of subhalo catalogues (as well as well-resolved merger trees in the case of SAMs), which can be challenging to obtain in $\sim$Gpc$^3$ volume simulations due to the large memory and CPU requirements needed to output them. Recent advances in HOD modelling \citep[see e.g.,][]{2020MNRAS.493.5506H,2021MNRAS.501.1603H,2021MNRAS.502.3242X,2021MNRAS.507.4879X,2022MNRAS.510.3301Y} allowing for greater flexibility and accuracy of the model have recast the HOD method into a favored choice for efficiently populating cosmological volume simulations with galaxies in the era of large redshift surveys. 

Here, we employ the \textsc{AbacusSummit} suite of high-performance cosmological $N$-body simulations \citep{2021MNRAS.508.4017M}, designed to meet the simulation requirements of the Dark Energy Spectroscopic Instrument (DESI) survey and run with the high-accuracy cosmological code \textsc{Abacus} \citep{2018ApJS..236...43G,2019MNRAS.485.3370G,2021MNRAS.508..575G}. In particular, we populate its halo light cone catalogues, which have been shown to be accurate at the sub-percent level \citep{2022MNRAS.509.2194H} with galaxies according to the \textsc{AbacusHOD} model, which equips the baseline HOD model with various generalizations pertaining to halo-scale physics and assembly bias \citep{2022MNRAS.510.3301Y}.

In this paper, we describe the \textsc{AbacusSummit} weak lensing products, which consist of octant and all-sky lensing maps with sub-arcmin resolution as well as source and lens galaxy mock catalogues necessary for simulating cross-correlations between the galaxy and lensing fields. We detail the construction process of the convergence, shear, and deflection field maps as well as the assignment procedure of various synthetic lensing properties to the HOD galaxy mock samples. We validate these observables using basic lensing statistics, such as the convergence angular power spectrum and the shear 2-point correlation function, and the cross-correlations of foreground and background galaxy samples. Several similar efforts for generating synthetic weak lensing observables exist in the literature. The Buzzard suite of weak lensing mocks \citep{2019arXiv190102401D} are similar to our work, but instead of the Born approximation, they adopt a ray-tracing technique, which allows one to use lower-resolution simulations. We note that the Buzzard mocks cover a smaller area on the sky and have lower resolution than \textsc{AbacusSummit}, though the ADDGALS technique allows for the injection of a high-density galaxy sample at arbitrary redshifts. Another notable effort in the realm of synthetic lensing catalogues is the CosmoDC2 suite of mocks \citep{2019ApJS..245...26K}, which is run on a single, but larger box and employs a ray-tracing recipe (output at $N_{\rm side} = 4096$). As a result, the generated patches are smaller but probe deeper redshifts ($z \approx 3$). A benefit of these mocks is that in addition to the lensing properties, they also output stellar mass, morphology, spectral energy distributions, and broadband filter magnitudes via a semi-analytic prescription. We also highlight the Stage-III-oriented CosmoGridV1 synthetic maps \citep{2022arXiv220904662K}, which are applied to a variety of cosmologies and adopt the Born approximation. Due to the computational expense, they utilize lower-resolution and smaller simulations, and address box repetition issues via a novel shell permutation scheme. Planned for the near future is the generation of \textsc{AbacusSummit} weak lensing catalogues on the extended cosmology grid, which similarly to the CosmoGridV1 mocks, will be enhanced by baryonification recipes and used to constrain cosmological parameters from Stage-IV survey data. Recently, the MillenniumTNG team released high-fidelity weak lensing maps generated from a full-physics hydrodynamical simulation adopting the Born approximation \citep{2023arXiv230412338F}. Other simulations focusing on weak-lensing statistics derived via full ray-tracing include \citet{2017ApJ...850...24T,2018MNRAS.481.1337H}.

This paper is organized as follows: Section \ref{sec:sim} briefly introduces the \textsc{AbacusSummit} suite of simulations and accompanying products relevant to this study. In Section \ref{sec:lens} we explain our procedure for generating weak lensing maps from the dark-matter outputs on the light cone, and validate the products. Section \ref{sec:mock} details and tests our recipe for assigning lensing properties to mock galaxies by using 2-point shear auto and cross-correlation statistics in harmonic and configuration space. In Section \ref{sec:obs}, we describe our pipeline for generating realistic DESI-, DES-, HSC- and KiDS-like mock catalogues. Finally, in Section \ref{sec:conc}, we summarize our main results and conclusions.

\section{Simulations}
\label{sec:sim}

In this Section, we introduce the \textsc{AbacusSummit} suite of high-performance cosmological $N$-body simulations and its relevant components, which were employed in the generation of the \textsc{AbacusSummit} weak lensing maps and catalogues. \textsc{AbacusSummit} \citep{Maksimova+2021} was designed to meet and exceed the Cosmological Simulation Requirements of the DESI survey. The simulations were run with \textsc{Abacus} \citep{2019MNRAS.485.3370G,Garrison+2021b}, a high-accuracy cosmological $N$-body simulation code, optimized for GPU architectures and  for large-volume simulations, on the Summit supercomputer at the Oak Ridge Leadership Computing Facility. 

The majority of the \textsc{AbacusSummit} simulations are made up of the so-called \texttt{base} resolution boxes, which house 6912$^3$ particles in a 2 Gpc$/h$ box, each with a mass of $M_{\rm part} = 2.1 \ 10^9\msunoh$. Additionally, we utilize the \texttt{huge} boxes with corresponding dimensions of 8640$^3$ particles in a 7.5 Gpc$/h$ box (with particle mass of $M_{\rm part} = 5.6 \ 10^{10}\msunoh$). While the \textsc{AbacusSummit} suite spans a wide range of cosmologies, here we focus on the fiducial outputs ($\Omega_b h^2 = 0.02237$, $\Omega_c h^2 = 0.12$, $h = 0.6736$, $10^9 A_s = 2.0830$, $n_s = 0.9649$, $w_0 = -1$, $w_a = 0$), consisting of 25 \texttt{base} (\texttt{AbacusSummit\_base\_c000\_ph\{000-024\}}) and two \texttt{huge} runs (\texttt{AbacusSummit\_huge\_c000\_ph\{201,202\}}). For full details on all data products, see \citet{Maksimova+2021}.

\subsection{Particle light cone catalogues}
\label{sec:pclelc}

\textsc{AbacusSummit} outputs a number of light cone quantities.
At every timestep, \textsc{Abacus} identifies particles that belong to the light cone and outputs their positions, velocities, particle IDs, and HEALPix pixel number. The particle outputs contain only a 10\% subsample of the particles, the so-called \texttt{A} and \texttt{B} subsamples, whereas the HEALPix products contain all particles. The pixel orientation of the latter is such that the $+z$ direction coincides with the North Pole. These maps come at a resolution of $N_{\rm side} = 16384$, corresponding to $\sim$0.2 arcmin, sufficient to probe the subhalo regime.

\begin{figure}
    \centering
    \includegraphics[width=0.48\textwidth]{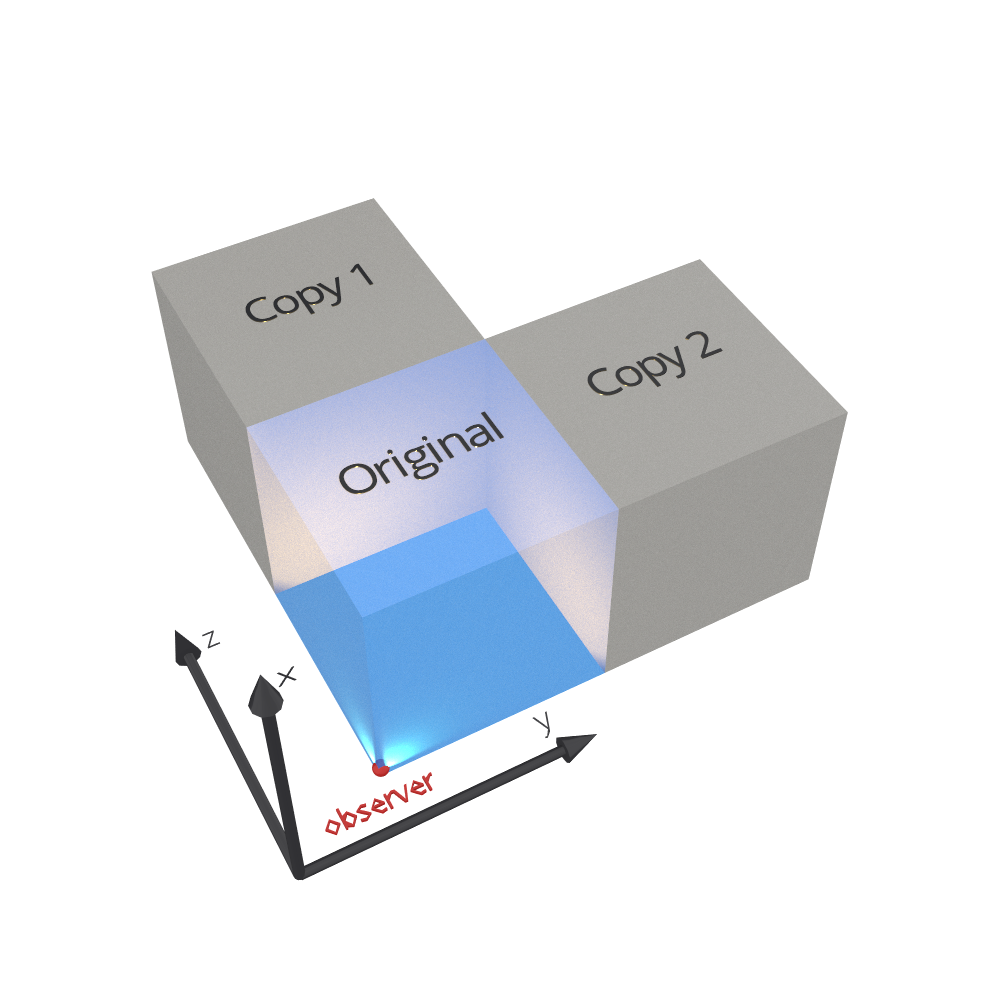} 
    \caption{Visualization of the geometrical arrangement of the \textsc{AbacusSummit} light cones. The original box, of length $2 \gpcoh$, is centered at (0, 0, 0), while two identical copies are placed at (0, 0, 2000) and (0, 2000, 0) $\mpcoh$. The observer is located at the corner of the original box, at (-990, -990, -990) $\mpcoh$.} 
    \label{fig:geo}
\end{figure}

The geometrical arrangement of the \texttt{base} light cones is shown in Fig. \ref{fig:geo}. In these simulations, which have box length of $2 \gpcoh$ on a side, the light cone observer is positioned at (-990, -990, -990), or, in other words, $10\mpcoh$ inside the corner of the original box. Three boxes, seemlessly attached to each other in an L shape, form the eligible space of the light cone, centered at (0, 0, 0), (0, 0, 2000), and (0, 2000, 0), respectively (measured in $\mpcoh$ units). Particles are output on-the-fly from every time step, where their trajectories are linearly interpolated to find the time when the light cone intersects their paths. Their positions and velocities are updated to this time. This provides an octant to a distance of $1990\mpcoh$ ($z \approx 0.8$), shrinking to two patches each about 900 square degrees at a distance of $3990\mpcoh$ ($z \approx 2.45$). We stress that all structure below $z \approx 0.8$ comes from the original box and is thus `unique,' while at higher redshifts, the repeated (overlapping) volume between the copies is minimal. In the case of the \texttt{huge} box simulations, the light cone is simply one copy of the box, centered at (0, 0, 0), providing a full-sky light cone to the half-distance of the box (3.75 Gpc/$h$, $z \approx 2.18$), which shrinks as one pushes further toward the eight corners (e.g., half the sky at $z = 3.2$).

\subsection{Halo light cone catalogues}
\label{sec:halolc}
The halo light cones of \textssc{AbacusSummit} provide an accurate and efficient mechanism for obtaining realistic galaxy catalogues on the sky via simulations, which is crucial for testing out systematic and observational effects, and developing pipelines for novel summary statistics. 

The procedure for obtaining the \textsc{AbacusSummit} halo catalogues, introduced in \citet{2022MNRAS.509.2194H}, starts with the \textsc{AbacusSummit} merger trees \citep{2022MNRAS.512..837B}, from which we calculate the interpolated positions and velocities of each halo at the redshift at which they cross the observer’s lightcone. We then associate the particles belonging to these haloes with the particle light cone outputs and report a number of new halo properties, in addition to the standard ones, such as merger-tree-interpolated and particle-averaged positions and velocities as well as interpolated masses. The thus obtained halo catalogues can then be populated with galaxies using the \textsc{AbacusHOD} prescription, which we describe next. We recommend using the halo light cone products in the halo mass regime of $M_{\rm halo} \gtrsim 2 \times 10^{11}\msunoh$, corresponding to haloes containing $\sim$100 particles or more, as at these mass scales the merger tree information is available to the vast majority of haloes and the particle-averaged quantities are less noisy. The \texttt{base} catalogues are generated for the redshift epochs: $z = 0.1, \ 0.15, \ 0.2, \ 0.25, \ 0.3, \ 0.35, \ 0.4, \ 0.45, \ 0.5, \ 0.575, \ 0.65, \ 0.725, \\ \ 0.8, \ 0.875, \ 0.95, \ 1.025, \ 1.1, \ 1.175, \ 1.25, \ 1.325, \ 1.4, \ 1.475, \ 1.55, \\ \ 1.625, \ 1.7, \ 1.775, \ 1.85, \ 1.925, \ 2.0, \ 2.25, \ 2.5$. The \texttt{huge} catalogues are available for all the same epochs until $z = 2.25$. The \textsc{AbacusSummit} halo light cone catalogues are available at DOI:\href{https://www.doi.org/10.13139/OLCF/1825069}{10.13139/OLCF/1825069}. 

\subsection{AbacusHOD}
\label{sec:hod}

The \textsc{AbacusSummit} halo light cone catalogues are designed to generate mock catalogues via \textsc{AbacusHOD}, a sophisticated routine that builds upon the baseline HOD model by incorporating various extensions affecting both the so-called one- and two-halo terms.

\textsc{AbacusHOD} allows the user to specify different tracer types: emission-line galaxies (ELGs), luminous red galaxies (LRGs), and quasistellar objects (QSOs). The \textsc{AbacusHOD} model is described in detail in \citet{2022MNRAS.510.3301Y}. 

\boryana{The \textsc{AbacusHOD} model incorporates several additional decorations beyond the vanilla HOD prescription, which are listed below:}
\begin{itemize}
    \item \texttt{s}  is the satellite profile modulation parameter, which modulates how the radial distribution of satellite galaxies within haloes deviate from the radial profile of the halo.
    \item \texttt{s\_v} is the satellite velocity bias parameter, which modulates how the satellite galaxy peculiar velocity deviates from that of the local dark matter particle.
    \item \texttt{alpha\_c} is the central velocity bias parameter, which modulates the peculiar velocity of the central galaxy.
    \item \texttt{s\_p} is the perihelion distance modulation parameter.
    \item \texttt{A\_c} or \texttt{A\_s} are the concentration assembly bias parameters for centrals and satellites, respectively.
    \item \texttt{B\_c} or \texttt{B\_s} are the environment assembly bias parameters for centrals and satellites, respectively. To define halo environment, we adopt the same formalism as \citet{2020MNRAS.493.5506H}.
    \item Parameterers specifying the central-satellite and satellite-satellite conformity of the ELGs, which has been found to affect non-trivially the one-halo term behaviour of star-forming galaxies \citep[see e.g.,][]{2019MNRAS.490.3532J,2022arXiv221010068H}.
\end{itemize}
We note that the assembly bias implementation preserves the overall galaxy number density by reranking haloes based on their pseudo-mass.

To emulate the intrinsic redshift-dependent change in the galaxy population, we implement a simple redshift-dependent HOD, which linearly (in scale factor, $a$) interpolates between the bestfit HOD values at a pair of redshifts for each tracer. The redshift-evolved value of parameter $\mu$ takes the following form (for a list of the parameters for each tracer, see Table~\ref{tab:hod}):
\begin{equation}
\label{eq:pivot}
    \mu_i(z) = \mu_{i, 0} + \mu_{i, p} \left(\frac{1}{1+z} - \frac{1}{1+z_{\rm pivot}}\right) ,
\end{equation}
where $\mu_{i, 0}$ and $\mu_{i, p}$ are the parameter value and its derivative computed at some redshift given two reference points, and $i$ iterates over all free parameters in the model. The values of the HOD parameters we adopt are discussed next.

\subsection{Dark Energy Spectroscopic Instrument}

The main propellant of this work is the need to create accurate weak lensing catalogues in anticipation of planned joint studies between the DESI redshift survey and photometric surveys that measure the induced shapes of galaxies by the effect of gravitational lensing.

DESI is a Stage IV dark energy experiment currently conducting a five-year survey of about a third of the sky with the goal to amass spectra for approximately 40 million galaxies and quasars \citep{2016arXiv161100036D}. The instrument operates on the Mayall 4-meter telescope at Kitt Peak National Observatory \citep{2022arXiv220510939A} and can obtain simultaneous spectra of almost 5000 objects over a $\sim$3$^{\circ}$ field \citep{2016arXiv161100037D,2022arXiv220509014S} thanks to a robotic, fiber-fed, highly multiplexed spectroscopic surveyor. The goal of the experiment is to unravel the nature of dark energy through precise measurements of the expansion history \citep{2013arXiv1308.0847L} and thus the dark energy equation of state parameters $w_0$ and $w_a$, with a predicted factor of five to ten improvement on their error relative to previous Stage-III experiments \citep{2016arXiv161100036D}.

\subsection{Galaxy samples}
\label{sec:galaxy}

\subsubsection{DESI}
To obtain the HOD parameters of our DESI-like LRG and ELG samples, we fit the predicted $\xi(r_p, \pi)$ correlations 
to the measured full-shape clustering on small scales ($r \lesssim 30\mpcoh$) from the DESI SV3 (Survey Validation 3) data. The fits are performed using the dynamic nested sampling package, \texttt{dynesty}, on the cubic \texttt{base}-resolution boxes. \texttt{dynesty} chains are run separately for each tracer and redshift epoch with covariance matrices computed from the $\sim$1800 \texttt{small} \textsc{AbacusSummit} boxes, designed with the express purpose of calculating covariance matrices for different summary statistics. The parameter values we arrive at are shown in Table~\ref{tab:hod} \citep{Yuan+2023}. 

\begin{table*}
\scriptsize
\centering
\begin{tabular}{||c c c c c c c c c c c c c c c c c c ||} 
 \hline
Tracer & $z$ & $p_{\rm max}$ & $\log(M_{\rm cut})$ & $\kappa$ & $\sigma$ & $\log(M_1)$ & $\alpha$ & $\gamma$ & $\alpha_c$ & $\alpha_s$ & $A_s$ & $\delta M_1$ & $\alpha_1$ & $\beta$ & $B_{\rm cen}$ & $B_{\rm sat}$ & $s$ \\ [0.5ex] 
\hline\hline
ELG & 1.1 & 0.156 & 11.65 & 8.98 & 2.47 & 15.32 & 1.29 & 5.08 & 0.0173 & 0.20 & 0.11 & -1.70 & -2.99 & 9.94 & -- & -- & -- \\
ELG & 0.8 & 0.116 & 11.52 & 3.58 & 1.18 & 15.97 & 1.29 & 4.04 & 0.180 & 0.523 & 0.289 & -2.54 & -2.59 & 6.77 & -- & -- & --\\
\hline
LRG & 0.8 & -- & 12.65 & 0.246 & 0.081 & 14.00 & 1.19 & -- & 0.165 & 0.943 & -- & -- & -- & -- & 0.117 & -0.921 & 0.124 \\
LRG & 0.5 & -- & 12.77 & 0.396 & 0.060 & 13.92 & 1.36 & -- & 0.308 & 0.913 & -- & -- & -- & -- & 0.063 & -0.206 & 0.481 \\ [1ex]
 \hline\hline
\end{tabular}
\caption{\textsc{AbacusHOD} parameters adopted for the DESI GQC galaxy mock catalogues. The redshift-evolving HOD parameters of the lensing mocks are obtained by linearly interpolating the parameters in the table between $z = 0.3$ and $z = 1.4$. Parameters come from the best fits to the DESI early data release.}
\label{tab:hod}
\end{table*}

In the case of our lensing mocks, we take the best fit values at redshifts $z = \{0.5, \ 0.8\}$ for the LRGs and $z = \{0.8, \ 1.1\}$ for the ELGs without perturbing their values. Assuming that these parameters evolve linearly with the scale factor $a$, we then calculate the parameter values at each redshift epoch of interest. The minimum and maximum values each parameter can take are set by the prior used in the fitting (see \citet{Yuan+2023}). That way we avoid adopting unphysical values for our HOD. 
The redshifts over which the HODs are applied to both LRGs and ELGs span between $z = 0.3$ and $z = 1.4$. These are chosen to roughly cover the range for which the target selection of the tracers is considered robust and the relative number density of the galaxies is large.

\subsubsection{HSC, DES, KiDS}
For the mock galaxy catalogues of the weak lensing surveys, rather than applying an HOD, we make use of the \textsc{AbacusSummit} halo light cone catalogues, selecting all haloes above a minimum mass cut to match the required source number density at each redshift (see Sec.\ref{sec:redsh} for more details). The reason we opt for this rather simplistic scheme is that in this study, we are not interested in modeling the clustering of the sources but rather only their cross-correlation with the lens sample, which is largely insensitive to the particular choice of a source population model. \boryana{We note that since we only use haloes for our sources, the source mock catalogues as such are unsuitable for estimating highly non-linear-scale effects such as boost factors and clustering redshifts. However, generating an alternative source galaxy sample on the light cone using AbacusHOD and equipping it with weak lensing properties (i.e., shears and deflections) is trivial, and our pipeline and products provide the necessary machinery for doing that.}

\section{Weak lensing maps}
\label{sec:lens}

Matching the halo light cone catalogues, we generate weak lensing maps for the 25 \texttt{base} boxes ($\texttt{AbacusSummit\_base\_c000\_ph\{000-024\}}$) and the two \texttt{huge} boxes ($\texttt{AbacusSummit\_huge\_c000\_ph\{201,202\}}$) at the fiducial \textsc{AbacusSummit} cosmology. We retain the native resolution of the HEALPix shell outputs on the light cone, i.e. $N_{\rm side} = 16384$, corresponding to a pixel size of 0.21 arcmin. We note that the pixel size is thus larger than the force softening scale of \textsc{AbacusSummit}, 7.2 kpc$/h$ (proper), at all available redshifts ($z > 0.15$), so we should not expect the lensing observables to be affected.

As detailed in this Section, we compute the lensing observables by weighting the shells by the appropriate lensing kernels, adopting the so-called ``Onion Universe'' approach \citep{2015MNRAS.447.1319F}, \boryana{which assumes the Born approximation}. We note that the particle counts shells are generated on-the-fly for each \textsc{Abacus} time step, of which there are $\gtrsim$1000 per simulation. Thus, these shells are output with excellent time granularity of $\Delta \log(a) \sim 0.001$ ($\lesssim 10\mpcoh$) at the relevant redshifts, so we expect the Born approximation to hold with sufficient accuracy for our configuration. Maps of the lensing observables -- convergence, shear, and deflection field, are available at 48 source redshifts, covering a redshift range of $z = 0.15$ to $z = 2.4$ ($\Delta z = 0.05$) as well as at the redshift of recombination, $z_{\rm rec} \approx 1089.3$. The geometry of the weak lensing maps is the same as the rest of the light cone products (see Fig.~\ref{fig:geo} and the discussion in Section \ref{sec:pclelc}). To perform the harmonic space operations, we employ the \texttt{healpy} and \texttt{ducc} packages. Comparisons with theory are accomplished via the Cosmological Core Library, \texttt{pyccl} package.

All products are publicly available under \href{https://app.globus.org/file-manager?origin_id=fc0006ee-68fc-11ed-8fd2-e9cb7c15c7d2&origin_path=%2F}{this URL}.

\subsection{Convergence}
\label{sec:conv}

\begin{figure}
    \centering
    \includegraphics[width=0.48\textwidth]{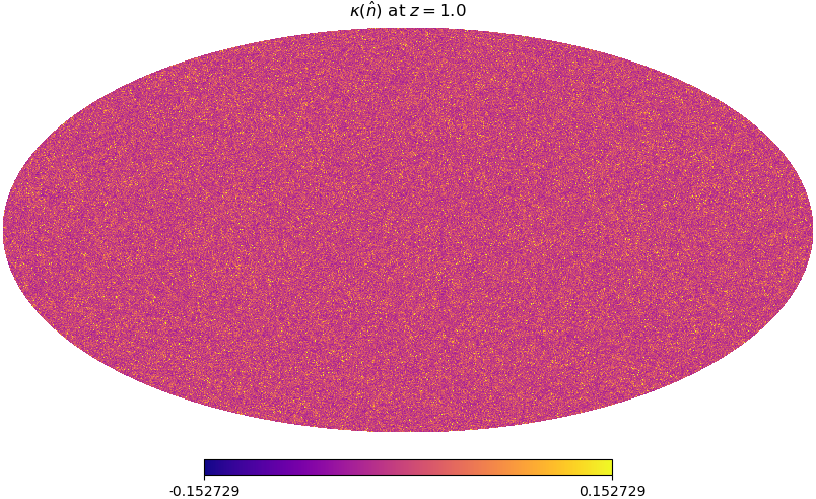}
    \includegraphics[width=0.48\textwidth]{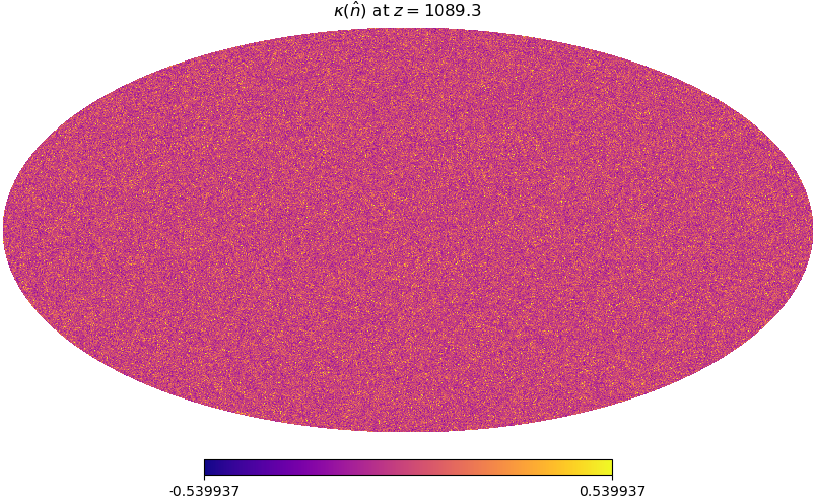}
    \caption{Convergence field at source redshift, $z_s$, of 1 (top panel) and 1089.3 (bottom panel), corresponding to the epoch of recombination. Maps shown for the \texttt{AbacusSummit\_huge\_c000\_ph201} simulation, which places the observer at (0, 0, 0) and thus covers the full sky until $z = 2.18$.}
    \label{fig:kappa}
\end{figure}

To obtain the convergence maps from the particle light cones, we adopt the Born approximation. Assuming flat space, the convergence field of the lensing distortions is given by
\begin{equation}
\kappa(\theta) = {\frac{3 H_0^2 \Omega_m}{2c^2}}~\int ~d \chi~\delta(\chi,\theta)
\frac{(\chi_s-\chi)\chi}{\chi_s~a}
\label{eq:kappa}
\end{equation}
where $H_0 = 100 h$ km/s/Mpc is the Hubble constant, $\Omega_m$ is the energy density of matter, $c$ is the speed of light,  $\delta$ is the three-dimensional matter overdensity at
radial distance $\chi(a)$, $a$ is the scale factor, $\theta$ is the angular position, and $\chi_s$ is the
distance to the lensing source(s). \boryana{Throughout the paper, we use comoving distances in units of $\mpcoh$.}

In order to compute the convergence map in an $N$-body simulation by adding up the light cone shells weighted by the appropriate weak-lensing kernel at each redshift, we can discretize this equation as
follows \citep{2008MNRAS.391..435F}:
\begin{equation}
    \kappa(i) = \frac{3H_0^2\Omega_{cb}}{2c^2}~\sum_j ~\delta(ij)~\frac{(\chi_s-\chi_j)\chi_j}{\chi_s a_j}~d\chi_j 
\end{equation}
where $i$ indicates the pixel position in the sky and $j$ the radial bin index \boryana{(i.e. distance to the mid-point $\chi_j$\footnote{Note that the field reported in the \textsc{AbacusSummit} light cone files, \texttt{CoordinateDistanceHMpc}, refers to the distance to the outer edge of the shell.} and the width $d\chi_j$ of the shell $j$)} into which we have sliced the simulation. Note that since \textsc{AbacusSummit} uses a basic prescription for neutrinos, modeling them as a smooth, non-clustering matter component \citep{Maksimova+2021}, we need to consider the contribution only from the gravitational components, i.e. baryons and cold dark matter, $\Omega_{cb} = \Omega_b + \Omega_{c}$. Such a treatment of neutrinos captures accurately the suppression on small scales, but does not account for the neutrino clustering on large scales. However, this is a secondary effect and does not matter for most applications relevant for galaxy surveys. 

We can compute the overdensity as:
\begin{equation}
    \delta(ij) = \frac{\rho(ij)}{{\bar{\rho}}}-1
\end{equation}
where $\bar{\rho}= \langle \rho(ij) \rangle$ is the mean density, which we compute analytically as $(N_{\rm part}/L_{\rm box}^3) {\Delta \Omega ~\chi_j^2~ d\chi_j}$, and the density per pixel is
\begin{equation}
\rho(ij) = \frac{N_{ij}}{{dV_j}} = \frac{N_{ij}}{{\Delta \Omega ~\chi_j^2~ d\chi_j}}
\end{equation}
where $\Delta\Omega$ is the area of each pixel and $N_{ij}$ is the number of particles in pixel $i$ and slice \boryana{(or equivalently, shell)} $j$. In Fig.~\ref{fig:kappa}, we show the convergence map at two source redshifts, $z_s = 1$ and 1089.3, for the \texttt{huge}-resolution simulation, \texttt{AbacusSummit\_huge\_c000\_201}, which covers the full sky.

The convergence field, $\kappa$, is related to the lensing potential through the two-dimensional Poisson equation,
\begin{equation}
\kappa(\hat{ \rm{n}}) = \nabla^2 \phi(\hat{\rm{n}}) 
\end{equation}
where $\phi (\hat{\rm{n}})$ is the lensing potential at a given point on sky, $\hat{\rm{n}}$. In harmonic space, this equation is greatly simplified, and the coefficients of the spherical harmonic transform, $\kappa(\hat{\rm{n}})=\sum_{\ell,m} \kappa_{\ell m} Y_{\ell m}(\hat{\rm{n}})$, are given by,
\begin{equation}
\kappa_{\ell m} = -\frac{1}{2}\ell(\ell+1) \phi_{\ell m} .
\end{equation}
Thus, at each source plane, we can compute the lensing potential from the convergence map by applying the appropriate $\ell$ weights in harmonic space. Similarly, thanks to their simple relations to the lensing potential in harmonic space, we can obtain other lensing observables.

\subsection{Deflection angle}
\label{sec:defl}
The deflection angle is related to the lensing potential through its gradient, \citep{2000PhRvD..62d3007H},
\begin{equation}
\alpha(\hat{\rm{n}}) = \nabla \phi(\hat{\rm{n}}) .
\end{equation}
In harmonic space, its coefficients are given by,
\begin{equation}
\alpha_{\ell m} = -\sqrt{\ell(\ell+1)} \phi_{\ell m} .
\end{equation}
Similarly, we can relate the deflection angle to the convergence field as
\begin{equation}
\alpha_{\ell m} = \frac{2}{\sqrt{\ell(\ell+1)}} \kappa_{\ell m} .
\end{equation}
The corresponding power spectra are thus simply related via \\ $4 \ \ell(\ell+1) \ C_{\ell}^{\alpha \alpha} = C_{\ell}^{\kappa \kappa}$. To obtain the real-space quantity, we transform back to real space. In \texttt{healpy} and \texttt{ducc}, the default $\ell_{\rm max}$ when performing the spherical harmonic transform is typically set as $\ell_{\rm max} = 4 \times N_{\rm side}$. However, at $N_{\rm side} = 16384$, this constitutes a huge computational expense,\footnote{Private correspondence with Martin Reinecke.} so instead we curb it $\ell_{\rm max} = 2 \times N_{\rm side}$ and check via lower-resolution maps that this does not affect the power spectrum of the lensing quantity measured at $\ell_{\rm max}$.

\subsection{Shear}
\label{sec:shear}
Finally, the shear maps, $\gamma (\hat{\rm{n}})$, can be related to the convergence field in spherical harmonic space \citep[see][]{2000PhRvD..62d3007H} via:
\begin{equation}
  \label{eq:shear}
\gamma_{\ell m} = - \sqrt{(\ell+2)(\ell-1)/(\ell(\ell+1))} \kappa_{\ell m} \equiv - f(\ell) \kappa_{\ell m} 
\end{equation}
and to the lensing potential via:
\begin{equation}
\gamma_{\ell m} = \frac{1}{2} f(\ell)
\ell(\ell+1) \phi_{\ell m} ,   
\end{equation}
with, $f(\ell)=\sqrt{(\ell+2)(\ell-1)/(\ell(\ell+1))}$.
Assuming that the B-mode signal is zero for the cosmological weak-lensing signal, the shear E-mode harmonic coefficients are given by $E_{\ell m}=\gamma_{\ell m}$, whereas the ``Stokes'' parameters of the shear field, $\gamma_1,\gamma_2$, can be obtained by transforming back the $E_{\ell m}$ coefficients to real space:
\begin{equation}
\gamma_1(\hat{\rm{n}}) \pm i\gamma_2(\hat{\rm{n}}) = \sum_{\ell m}
\gamma_{\ell m} Y_{\ell m}(\hat{\rm{n}})
\end{equation}
Analogously to the case of the deflection field, we set the maximum $\ell$-mode of that transformation to $\ell_{\rm max} = 2\times N_{\rm side}$.

\subsection{Validation}
\label{sec:lensvalid}

Having obtained the convergence maps, we can compare it with the theoretically estimated angular power spectrum from \texttt{pyccl} \citep{1812.05995} given by:
\begin{equation}
    C^{\kappa\kappa}_\ell(\kappa) = \frac{9H_0^4\Omega_m^2}{{4c^4}}~\int ~d\chi~ P(k,z) \frac{(\chi_s-\chi)^2}{{\chi_s^2~a^2}}
\end{equation}
where $P(k,z)$ is the three-dimensional density power spectrum at redshift $z$ evaluated at $k=\ell/\chi$ in the Limber approximation \citep{1953ApJ...117..134L}, which is known to be valid for $\ell>10$ within a few percent accuracy \citep[see e.g.,][]{2002PhR...367....1B}.  

\begin{figure}
    \centering
    \includegraphics[width=0.48\textwidth]{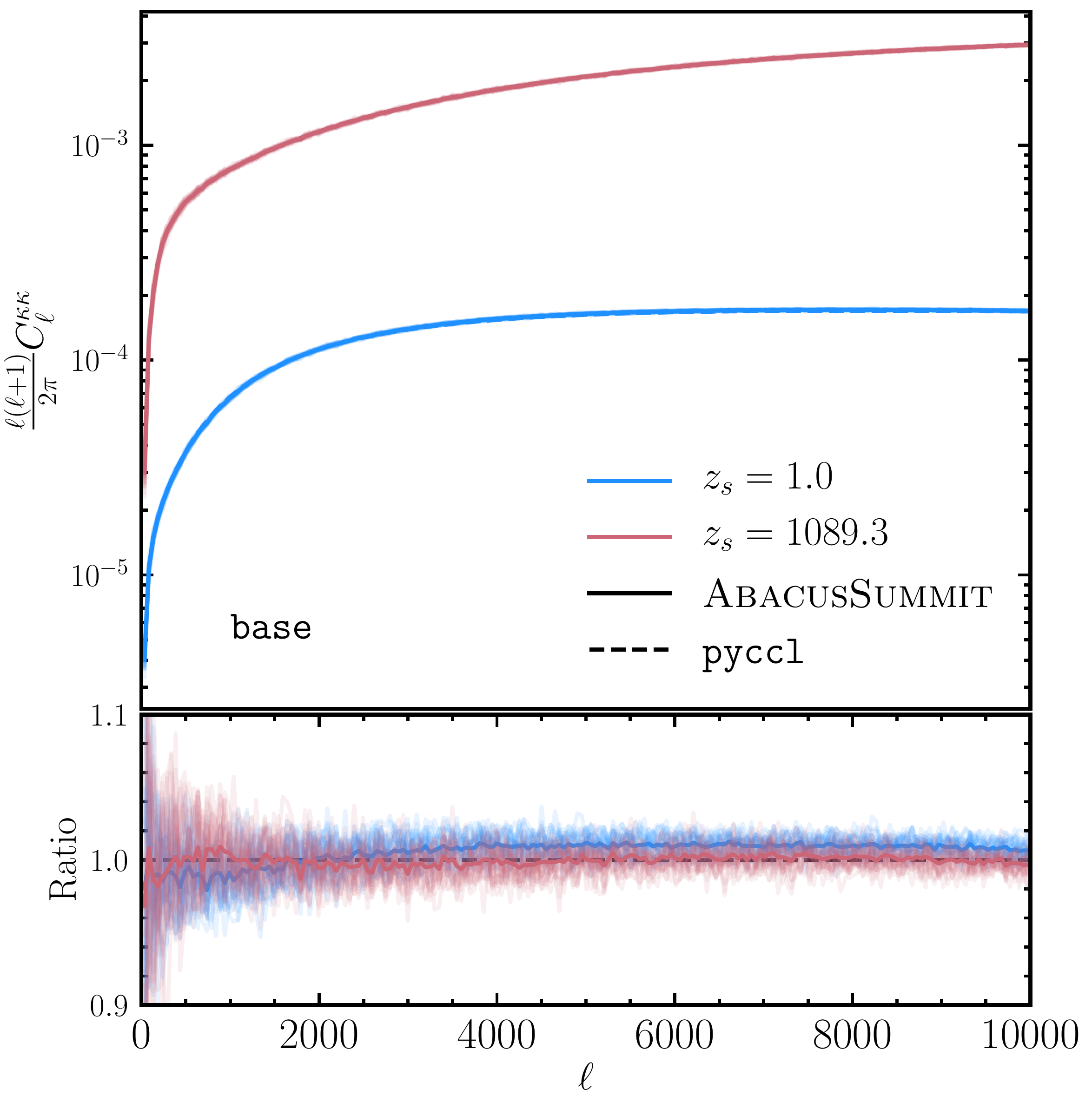}
    \includegraphics[width=0.48\textwidth]{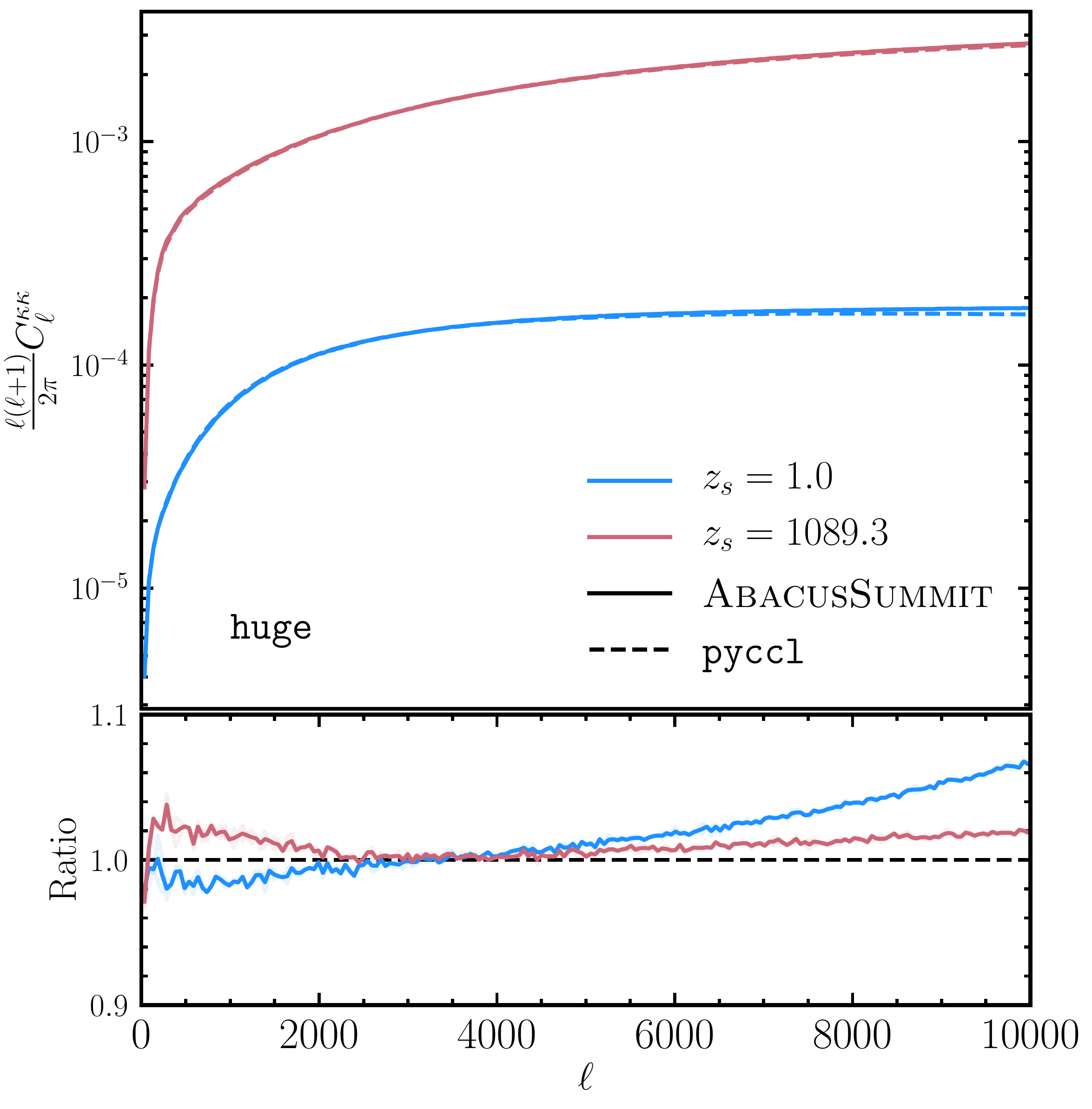}
    \caption{Convergence angular power spectrum at two source redshifts, $z_s = 1$ (blue) and $z = 1100$ (red), computed using \textsc{AbacusSummit} (solid) and the cosmological library \texttt{pyccl} (dashed). The top panel shows the results for the 25 base-resolution simulations (covering less than an octant), while the bottom is obtained using the two \texttt{huge}-resolution simulations (covering the full sky). Each panel also shows the ratio between the simulation and the theoretical prediction, which is calculated using \texttt{halofit} (total matter). In shaded colors, we plot the curves from all simulations. The simulation curves in the top panel demonstrate remarkable agreement with theory for all scales considered ($30 < \ell < 10000$). On smaller scales, cosmic variance contributes to the uncertainty in the measurement. For the \texttt{huge} simulations, the CMB lensing power spectrum deviates at the 2\% level with a mild scale dependence, while at $z_s = 1$, the discrepancy reaches 6\%. We attribute this difference to resolution effects and further study this in Appendix~\ref{app:halofit}.}
    \label{fig:clkk}
\end{figure}

Fig.~\ref{fig:clkk} shows the comparison between the convergence power spectrum measured from the \textsc{AbacusSummit} maps and the non-linear theoretical fit (i.e., \texttt{halofit}, as implemented in \texttt{pyccl}), for sources at $z_s = 1$ and $z_s = 1089.3$. We show the results both for the 25 base-resolution simulations (covering a bit less than an octant at $z_s = 1$ and about 1800 sq. deg at $z_s = 1089.3$) as well as the two \texttt{huge}-resolution simulations (covering the full sky). The theoretical prediction is calculated using \texttt{pyccl}, which in turn uses the \texttt{halofit} non-linear matter power spectrum. The base-resolution simulation curves demonstrate remarkable agreement with theory for all scales considered ($30 < \ell < 10000$). In particular, the CMB lensing agreement is $\lesssim$0.1\% on average, whereas for $z_s = 1$, it is about 1\%. We note that the quoted accuracy of \texttt{halofit} is about 5\% \citep{2012ApJ...761..152T}, so the agreement we observe is far below that threshold. We attribute the worse agreement at $z_s = 1$ between \texttt{pyccl} and \textsc{AbacusSummit} to the fact that \texttt{halofit} provides a poorer fit to the simulation power spectrum at low redshifts (see Fig.~\ref{fig:halofit} and the commentary in Appendix~\ref{app:halofit}). At the lowest multipoles, sample variance introduces large fluctuations in the measured power for a single realization, contributing as much as 10\% to the uncertainty in the measurement. 

In the case of the \texttt{huge} simulations (shown on the bottom of Fig.~\ref{fig:clkk}), we see that the convergence power spectrum is more discrepant with \texttt{pyccl}. In particular, while the CMB lensing power spectrum deviates at no more than 2\% for all scales shown, the deviation is much more evident in the $z_s  = 1$ case (and also displays some scale dependence), at the smallest scales reaching 6\%. We attribute this difference, relative to the base simulations, to resolution differences. As a separate test, we swap the \texttt{halofit} non-linear power spectrum in \texttt{pyccl} with Cosmic Emu \citep{2016ApJ...820..108H} and find that the \texttt{huge} simulations convergence power spectrum is much better matched. We conjecture that this is because Cosmic Emu employs a simulation suite with resolution characteristics that are very similar to the \textsc{AbacusSummit} \texttt{huge} boxes. Until $\ell < 9000$, the agreement with \texttt{halofit} is within the 5\% expected margin. Below $\ell \lesssim 10$, we advise treating the comparisons with caution, as the Limber approximation (adopted in the power spectrum calculation) breaks down \citep{2008PhRvD..78l3506L}.

\begin{figure}
    \centering
    \includegraphics[width=0.48\textwidth]{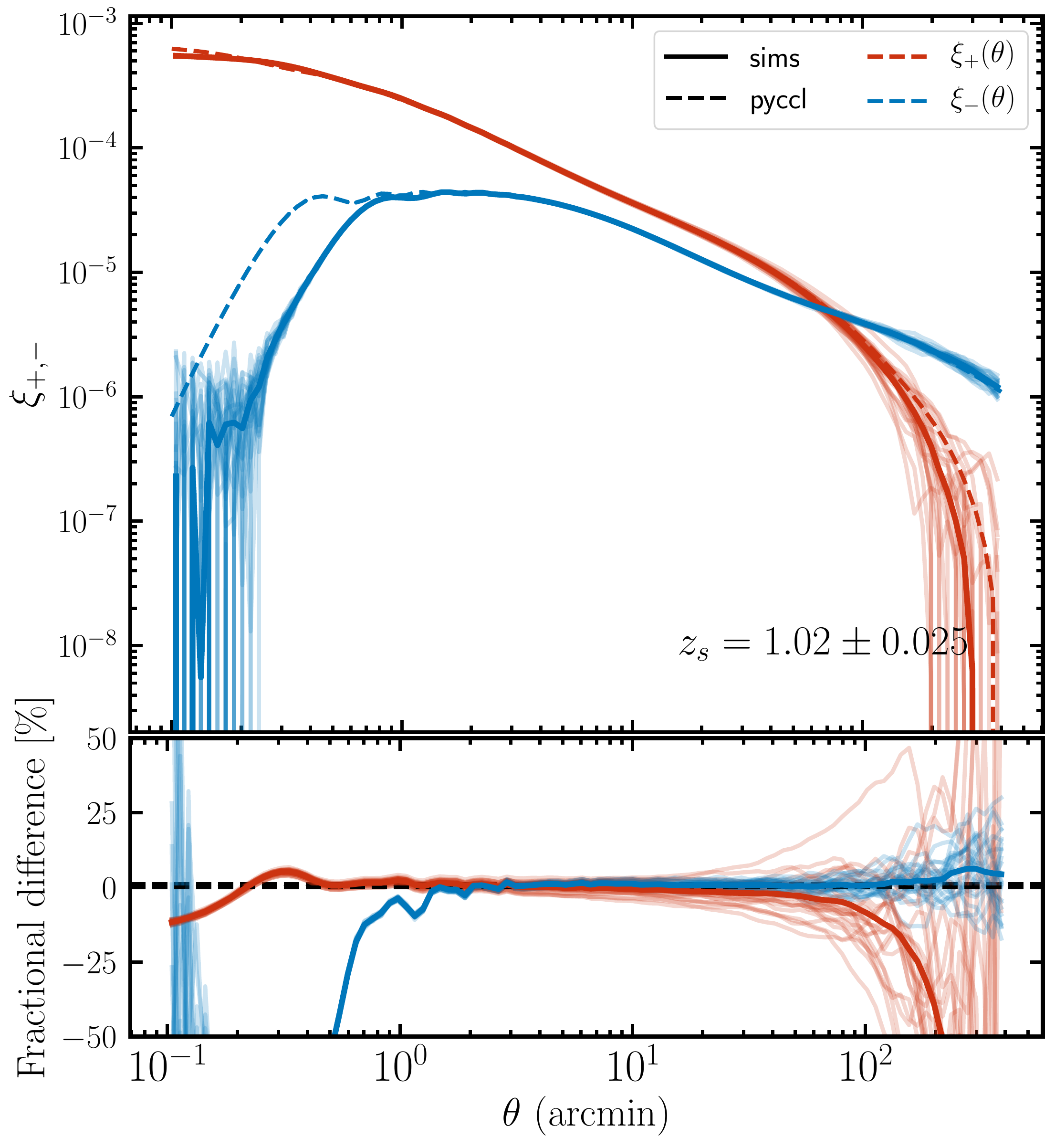}
    \caption{Shear-shear auto-correlation, $\xi_\pm (\theta)$, at $z_s = 1.025$ source redshift. Results are shown for all 25 base simulations (shaded), with the average $\xi_\pm$ displayed as solid curves (red and blue, respectively).The source catalogue consists of all haloes at that redshift epoch. We find that in the intermediate regime of $1' < \theta < 100'$, the agreement is within 5\%. \boryana{Below 0.7 arcmin, $\xi_-$ sees a stark cutoff due to
    resolution effects coming from the pixel size (0.2 arcmin).} 
    Above $\theta = 100$ arcmin, $\xi_+$ is more strongly affected by map boundary effects, which are investigated in detail in Appendix~\ref{app:shear}.}
    \label{fig:xipm}
\end{figure}

In Fig.~\ref{fig:xipm}, we show the shear-shear auto-correlation, $\xi_\pm (\theta)$, at source redshift of $z_s = 1$, computed for all 25 base-resolution simulations. The source catalogue is made up of all the haloes belonging to the source redshift epoch. We find that in the intermediate regime, i.e. $1' < \theta < 100'$, the agreement is within 5\%. Below 1 arcmin, $\xi_+ (\theta)$, which at fixed $\theta$ receives contributions from larger scales relative to $\xi_- (\theta)$ \citep{2002A&A...396....1S}, exhibits excellent agreement. \boryana{On the other hand, we see a steep cutoff in the $\xi_- (\theta)$ curve. This is the result of resolution effects due to the pixel size (0.2 arcmin), which affect $\xi_-(\theta)$ more noticeably than $\xi_+(\theta)$ as argued above. }
We note that the use of haloes rather than galaxies (particles + haloes) reduces the noise in the $\xi_\pm (\theta)$ measurement. 
Above $\theta > 100$ arcmin, $\xi_- (\theta)$ displays excellent agreement, whereas $\xi_+ (\theta)$ displays a stark cutoff. We conjecture that this is the result of boundary effects in the map construction procedure and investigate this idea in Appendix~\ref{app:shear} through several test on the full-sky \texttt{huge}-resolution box. We conclude that there are two main boundary effects: on one hand, the area covered by the source maps, which is a bit less than an octant at $z_s = 1$, displays significant sample variance at these scales; on the other hand, the conversion of partial-sky $\kappa (\hat n)$ into $\gamma_{1,2}(\hat n)$ introduces features at the boundary, which affect the large-scale measurement. We note that at higher redshifts, these effects are more pronounced ($z_s \approx 0.8$ maps cover an octant of the sky, but the area covered shrinks at higher redshifts). Although our correlation measurements exhibit a deficit beyond $\theta > 100$, the errors on them are still within the noise of the weak lensing surveys of interest (see Fig.~\ref{fig:xipm_desy3}, where we mimic DES Y3 with realistic noise) and so have minimal effect on cosmological parameter fits.

\section{Mock catalogues}
\label{sec:mock}

In this Section, we detail our procedure for augmenting our mock galaxies with lensing properties via the weak lensing maps discussed in Section \ref{sec:lens}.

\subsection{Lensing assignment}
\label{sec:lassign}
Mock galaxies are output using the \textsc{AbacusHOD} method, as discussed in Section \ref{sec:hod} and Section \ref{sec:galaxy}. To assign lensing properties to each of our mock galaxies, we adopt the following simple algorithm: 

\begin{enumerate}
\item{
For a given galaxy with coordinates on the light cone $(\hat{\rm{n}}, z)$, where $\hat{\rm{n}}$ denotes its angular position in the sky and $z$ its redshift, we find the corresponding closest source plane in redshift and the sky coordinates pixel of the pixels which the galaxy occupies (i.e., discretized pixel center coordinates)}
\item{
Then, we look up the value of the relevant lensing quantities for this pixel and redshift from our convergence and shear weak lensing maps, $\kappa, \ \gamma_1, \ \gamma_2$, as well as the deflection angle values at this location, $\alpha_\theta$ and $\alpha_\phi$.}
\item{
Finally, for each galaxy, we record the values of the weak lensing fields, as well as the perturbed sky coordinates due to the tiny deflection to the light ray caused by gravity (discussed in Section~\ref{sec:magbias}).}
\end{enumerate}

This simple implementation is sensitive to the pixel resolution used, which in our case is 0.21 arcmin ($N_{\rm side} = 16384$).  Consistently, we only expect to model lensing observables accurately down to $\sim 0.3$ arcmin scales, as we will discuss in detail in Section~\ref{sec:obs}. Another limitation of this procedure is that different galaxies that occupy the same pixel will have identical lensing properties. Given the small physical scale that our choice of $N_{\rm side}$ corresponds to, we expect that this will play a very minor role. 

\subsection{Magnification bias}
\label{sec:magbias}
The gravitational lensing induced by large-scale structures on background sources changes their observed number density and thus introduces an additional cross-correlation signal between background and foreground galaxy populations \citep{1998MNRAS.294L..18M,2001PhR...340..291B}. This effect, first measured in cross-correlations between distant quasars and low-redshift galaxies (e.g., \cite{1997ApJ...477...27B,2003ApJ...589...82G,2005MNRAS.359..741M,2005ApJ...633..589S}), is known as ``magnification bias'' and can be used to constrain the galaxy-mass power spectrum \citep{2003MNRAS.345...62J}.

To build theoretical intuition, let us consider the case of a magnitude limited survey. Then, the cumulative number of galaxies above a certain flux limit, $f$, is roughly proportional to $N_{0}(>f) \propto A f^{\alpha}$, with $A$ is the area of the survey and $\alpha $ the power-law slope of the background (source) number counts. Because lensing preserves the surface brightness of galaxies, the observed survey depth increases (i.e. the effective flux limit decreases), whereas the effective area decreases by the same amount:  $f \rightarrow f/\mu$, $A \rightarrow A/\mu$, where $\mu$ is the magnification. These two counteracting effects induce a bias in the cumulative number of background sources given by:
\begin{equation}
N(>f) \propto  \frac{1}{\mu} A \left(\frac{f}{\mu}\right)^{-\alpha} = \mu^{\alpha-1} N_{0}(>f) .
\label{eq:cumnumgal}
\end{equation}
In the weak-lensing limit, we can approximate the magnification as $\mu \approx 1 + \delta_{\mu}$ since $|\delta_{\mu}| \ll 1$. Taylor expanding the magnification term, $\mu^{\alpha-1} \approx 1 + (\alpha-1) \delta_{\mu}$, we arrive at the following expression for the additional overdensity contributed by the magnification,
\begin{eqnarray}
\delta_{\rm mb} &=& \frac{N-N_{0}}{N_{0}} = \delta_{\rm mm} + \delta_{\rm lp} \nonumber
\\
&=&  (\alpha-1)\delta_{\mu} =2(\alpha-1)\delta_{\kappa}
\label{eq:deltanumgal}
\end{eqnarray}
where the last equality uses the weak lensing approximation that relates the fluctuations in magnification and convergence via $\delta_{\mu} \approx 2\,\delta_{\kappa}$. Note that $\delta_{\rm mb}$ (``mb'' stands for magnification bias) incorporates two qualitatively different contributions:
\begin{enumerate}
\item{{\rm extra fluctuations from magnified magnitudes, $\delta_{\rm mm} = \alpha \,\delta_{\mu}$}}
\item{{\rm extra fluctuations from  lensed positions, $\delta_{\rm lp} =-\delta_{\mu}$}} .
\end{enumerate} 
While the two effects cannot be separated observationally, we implement them in our simulations using two distinct procedures, which are outlined next. We can thus test and verify each one of them separately. Traditionally, we work with the logarithmic slope of the background
number counts, $s$, at redshift $z$, for a magnitude limit $m$, defined as:
\begin{equation}
s = 2\alpha/5 \equiv \frac{ d {\rm Log_{10}}N(<m,z)}{dm}.
\end{equation} 
The overall contribution to the counts due to magnification bias thus becomes
\begin{equation}
    \delta_{\rm mb}=(5s - 2)\delta_{\kappa}.
\end{equation}

The net magnification effect depends on how the loss of sources due to the area contraction, $\delta_{\rm lp}$, is compensated by the gain of sources from the flux magnification, $\delta_{\rm mm}$. Number counts for source populations with flat luminosity functions, $s < 0.4$, such as faint galaxies, decrease due to the effective shrinkage in area, whereas sources with steep luminosity functions, $s > 0.4$, such as quasars, increase due to the effective deepening of the survey. In the case of $s=0.4$ ($\alpha=1$), there is no net change to the counts due to magnification bias.

\subsection{Implementation of magnification bias}
\label{sec:implmag}
In this Section, we describe how we implement magnification in the magnitudes and positions of mock galaxies. 
\begin{enumerate}
\item{{\bf magnified magnitudes:} flux magnification causes the mock galaxy magnitudes, $m$, to become brighter by the following amount:
\begin{equation}
\Delta m = 2.5 {\rm Log_{10}}\mu \approx \frac{5}{\ln{10}} ~\kappa
\label{eq:deltam}
\end{equation}
where we have Taylor expanded $ \mu \approx 1+\delta\mu$ and used the weak lensing relation $\delta\mu \approx 2\kappa$ in the weak-lensing limit. Knowing the value of the convergence, $\kappa$ at the location of a given galaxy, we can compute the induced magnification to its flux, which in turn leads to a change in the source number counts, $\delta_{\rm mm}$. In the case of the \textsc{AbacusHOD} model, we do not output magnitudes or fluxes but only galaxy positions and velocities. For this reason, we implement the effect of magnitude magnification by weighting the galaxies by:
\begin{equation}
    w = 1 + 2 \alpha \ \delta_\kappa = 1 + 5 s \  \delta_\kappa ,
\end{equation}
which effectively boosts the contribution of galaxies located at local peaks of the convergence field.
}
\item{{\bf lensed positions:} the observed position of a source galaxy, ${\pmb\beta}$, is shifted from its true (unlensed) position, ${\pmb\theta}$, by an angle determined by the deflection field vector, ${\pmb\alpha}$, at its true location (for details, see \citet{2001PhR...340..291B}). In the single-plane (i.e., Born) approximation, 
the lens equation reads,
\begin{equation}
{\pmb\beta} = {\pmb\theta} - {\pmb\alpha}
\label{eq:lenseq}
\end{equation} 
where ${\pmb\alpha}$ is a two-dimensional vector, and the lensed position ${\pmb\beta}$ is obtained by moving the galaxy position along a geodesic on the sphere in the
direction of the vector  by an amount (arc length) given by the size of the deflection angle, $\alpha$. Guided by this, we can remap the source galaxy positions due to gravitational lensing in our galaxy outputs.

If we denote the unlensed position on the sphere by
${\pmb\theta} = (\theta,\phi)$, then the lensed position, ${\pmb\beta} = (\theta^{\prime},\phi +
\Delta \phi)$, can be simply written down as:
\begin{eqnarray}
\cos\theta^{\prime} &=&
\cos\alpha\,\cos\theta-\sin\alpha\,\sin\theta\,\cos\delta \nonumber \\
\sin\Delta\phi &=& \sin\alpha\,\sin\delta/\sin\theta ,
\label{eq:remap}
\end{eqnarray} 
where we have used the spherical triangle identities \citep{2005PhRvD..71h3008L,2008ApJ...682....1D} and the deflection vector is projected on the spherical polar basis, $(\mathbf{e_{\theta}},\mathbf{e_{\phi}})$, as 
${\pmb\alpha}=\alpha_{\theta}\,\mathbf{e_{\theta}} +
\alpha_{\phi}\,\mathbf{e_{\phi}}=\alpha\,\cos\delta \,\mathbf{e_{\theta}} + \alpha\,\sin\delta\,
\mathbf{e_{\phi}}$, with $\delta$ being the angle between the deflection vector and the polar basis vector $\mathbf{e_{\theta}}$.}
\label{eq:lens}
\end{enumerate}

In summary, in order to enable magnification bias in the lens catalogues, we recommend calculating the lensed positions of galaxies according to Eq.~\ref{eq:lens} and weighting the galaxies by the quantity in Eq.~\ref{eq:remap}. We adopt the `lensed positions' for the source catalogue when switching on magnification bias effects, which also affects the sources. 

\subsection{Validation}
\label{sec:mockvalid}

\subsubsection{Harmonic space}

In weak lensing studies, we typically split the data into redshift bins. To obtain the theoretical measure of the angular cross-power spectrum between two fields $\delta_\mathrm{a}$, $\delta_\mathrm{b}$, measured in redshift bins $i,j$, we can adopt the following equation
\begin{equation}
 C_{\mathrm{ab}}^{ij}(\ell) = \int d\chi \frac{q_\mathrm{a}^i(\chi)q_\mathrm{b}^j(\chi)}{\chi^2}P_{\mathrm{ab}}(\ell/\chi,z(\chi)),
\label{eq:cab}
\end{equation}
where $P_{\mathrm{ab}}(k,z)$ is the three-dimensional cross-power spectrum of the fields at redshift $z$ and wavenumber $k$, $\chi(z)$ is the comoving distance, and $q_\mathrm{a,b}(\chi)$ are the weight functions, \citep{1992ApJ...388..272K,PhysRevD.70.043009}, and we have assumed the Limber approximation \citep{2017JCAP...05..014L}, $k = \ell/\chi$. For the galaxy density field, $\delta_\mathrm{g}$, corresponding to the foreground (lens) galaxy population, the weight function $q_g(\chi)$ is given by
\begin{equation}
\label{eq:qg}
q_\mathrm{g}^i(\chi) =\frac{n^i_{\mathrm{lens}}(z)  }{\bar{n}_{\mathrm{lens}}^i}\frac{dz}{d\chi}\,,
\end{equation}
where $n^i_{\mathrm{lens}}(z)$ is the redshift distribution of the sample in redshift bin $i$ and $\bar{n}_{\mathrm{lens}}^i$  is the average lens density.  For the convergence field, $\delta_\kappa$, corresponding to the background (source) galaxy population, the weight function $q_\kappa(\chi)$ can be calculated as:
 \begin{equation}
\label{eq:qkappa}
q_\kappa^{i}(\chi) = \frac{3 H_0^2 \Omega_m }{2 c^2}\frac{\chi}{a(\chi)}\int_\chi^{\chi_{\mathrm{max}}} \, d \chi' \frac{n_{\mathrm{source}}^{i}(z)\,
}{\bar{n}_{\mathrm{source}}^{i}}\frac{dz}{d\chi'} \frac{(\chi'-\chi)}{\chi'} \,,
\end{equation}
where $H_0$ and $\Omega_m$ denote the values of the present-day Hubble parameter and matter density and $\chi_{\mathrm{max}}$ is the maximum comoving distance of the source distribution. Analogously to the lens case, $n^i_{\mathrm{source}}(z)$ and  $\bar{n}^i_{\mathrm{source}}$ are the source redshift distribution and average density of sources in redshift bin $i$.  We describe the redshift distributions of our source and lens populations in Section \ref{sec:redsh}.

\begin{figure}
    \centering
    \includegraphics[width=0.48\textwidth]{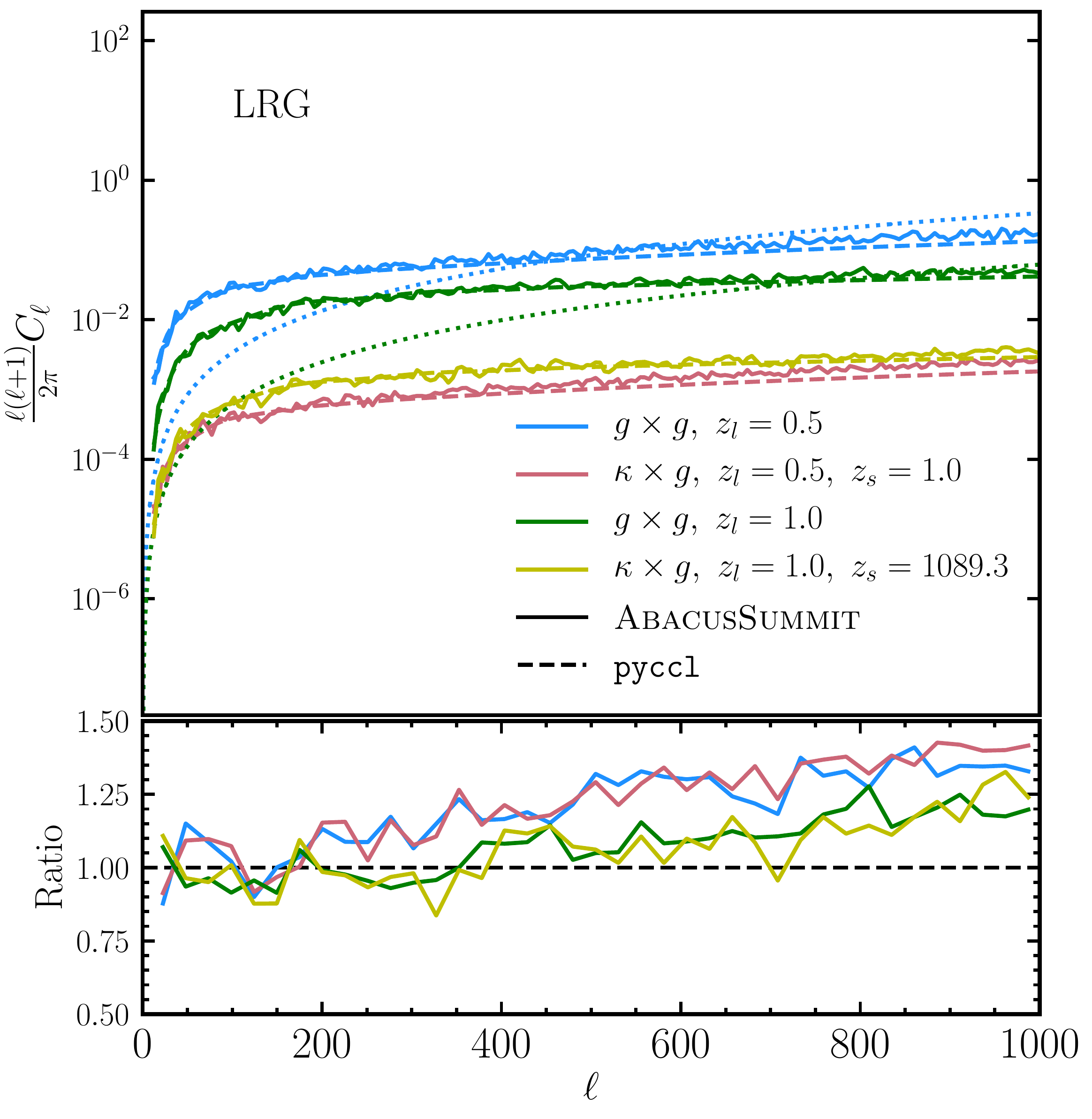}
    \includegraphics[width=0.48\textwidth]{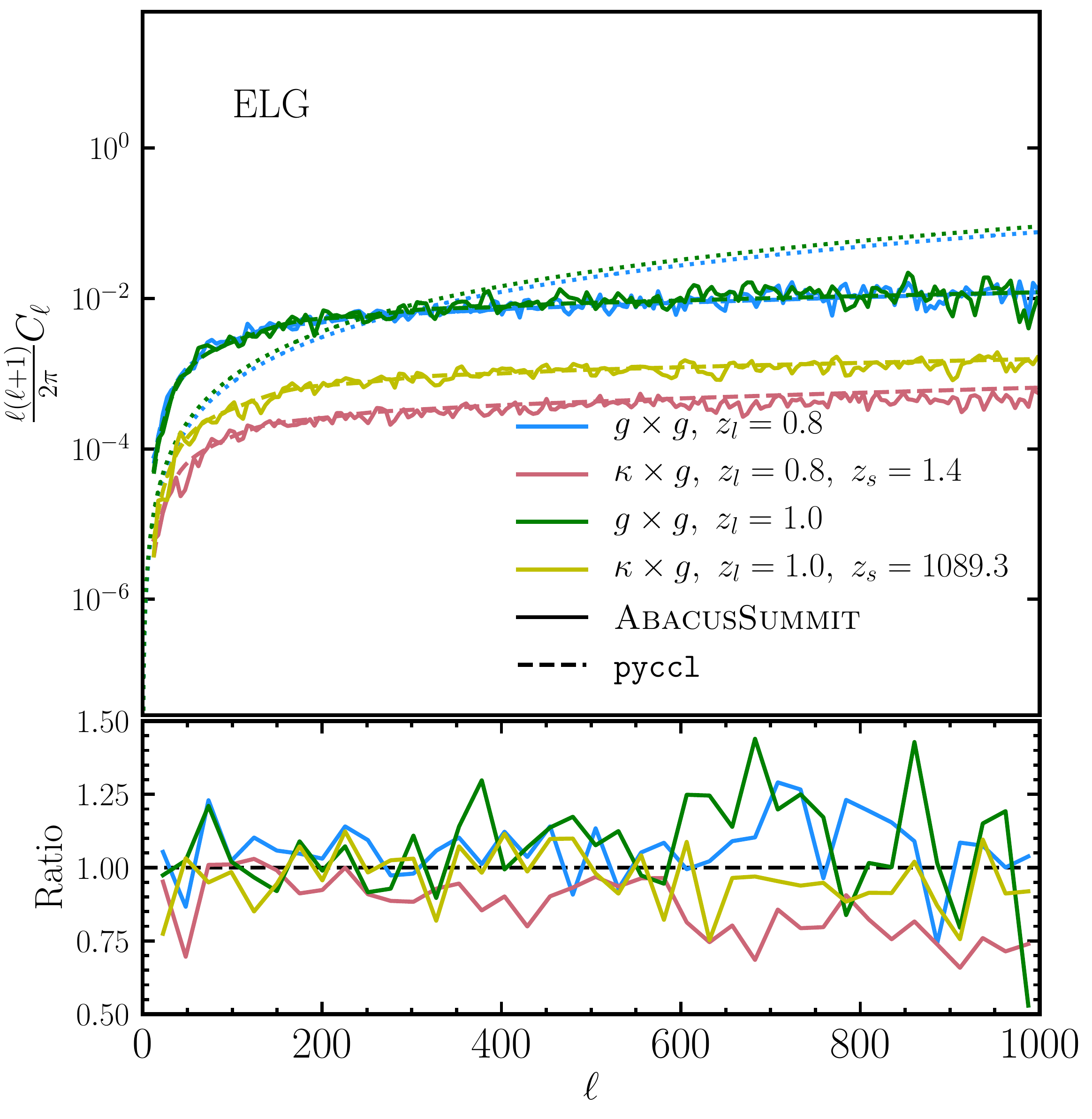}
    \caption{Angular power spectrum of the galaxy-galaxy and galaxy-convergence cross-correlation at two source redshifts $z_s = 1$ (1.4) and $z_s \approx 1100$ and two lens redshifts $z_l = 0.5$ (0.8) and $z_l = 1$ for LRGs (ELGs). The top panel corresponds to the LRG galaxy population, while the bottom corresponds to the ELG one. The dotted curves show the Poisson noise contribution to the signal, which we subtract from the $g \times g$ signal. The smallest scales we consider are $\ell < 1000$, as the linear bias approximation worsens, as we plunge into smaller and smaller scales. Similarly to Fig.~\ref{fig:wp} and Fig.~\ref{fig:gammat}, relative to the linear bias approximation, the small-scale clustering of ELGs is suppressed while that of LRGs is in excess. Results shown for \texttt{AbacusSummit\_base\_c000\_ph002}.}
    \label{fig:clkg}
\end{figure}

Fig.~\ref{fig:clkg} shows the angular power spectrum of the galaxy-galaxy and galaxy-convergence cross-correlation between two pairs of lens and source galaxy populations; namely, $z_l = 0.5$ ($z_l = 0.8$) and $z_s = 1$ ($z_s = 1.4$) as well as $z_l = 1$ with $z_s = 1100$ for LRGs (ELGs). We choose the pairings in a way that ensures the lens redshift distribution has good overlap with the lensing kernel of the source. We downsample the DESI lens populations of LRGs and ELGs (defined in Section~\ref{sec:galaxy}) to match a Gaussian form of the redshift distribution centered on $z = 0.5$ (0.8) and 1.025, respectively, with a width of $\Delta z = 0.2$. We subtract the Poisson noise contribution from the galaxy clustering signal, but note that the noise term may deviate from the Poisson assumption \citep[e.g.,][]{2022MNRAS.512.1829M}. The smallest scales we consider are $\ell = 1000$, as the linear bias approximation worsens as we venture far into the non-linear regime. 

At large scales, $\ell < 200$, we see that the agreeement with theory of all curves is excellent -- i.e., the ratio is consistent with one, within a 10\% error budget. Assuming that the scale at which the linear bias approximation breaks down is $k \approx 0.1\hompc$, which corresponds to $\ell \approx 180$ and $300$ for $z_l = 0.5$ and 1, respectively (using the Limber approximation), it is no surprise that we start seeing large deviations from the simple \texttt{halofit} + linear bias model around these scales. We note that as expected, the high redshift pair, which is affected by non-linear physics on smaller scales, follows well the \texttt{pyccl} prediction down to smaller scales. We choose the linear bias parameters by eye such that they fit the large scales of the galaxy auto-power spectrum. The values we arrive at are: 1.1 ($z = 0.8$) and 1.25 ($z = 1.025$) for the ELGs, and 1.95 ($z = 0.5$) and 2.3 ($z = 1.025$) for the LRGs. As will be discussed next in the next section (see Fig.~\ref{fig:wp} and Fig.~\ref{fig:gammat}), relative to the linear bias approximation, the small-scale clustering of ELGs is suppressed while that of LRGs is in excess. Due to the larger size of the LRG haloes, which in turn pushes the effects of the one-halo term to larger scales, it is no surprise that the deviations from the linear bias approximation kick in at smaller values of $\ell$. 
\boryana{We also observe in the lower (ratio) panels that for a given choice of tracer and redshift, the two curves, $\kappa \times g$ and $g \times g$, divided by their respective linear bias predictions, are in good agreement with each other. This indicates that the galaxy-matter cross-correlation coefficient is close to one. As expected, in the case of the ELGs, the ratios are less consistent with each other, as the cross-correlation coefficient exhibits a larger deviation from one \citep[see e.g.,][]{2022arXiv221010072H}.}

\subsubsection{Configuration space}

\begin{figure}
    \centering
    \includegraphics[width=0.44\textwidth]{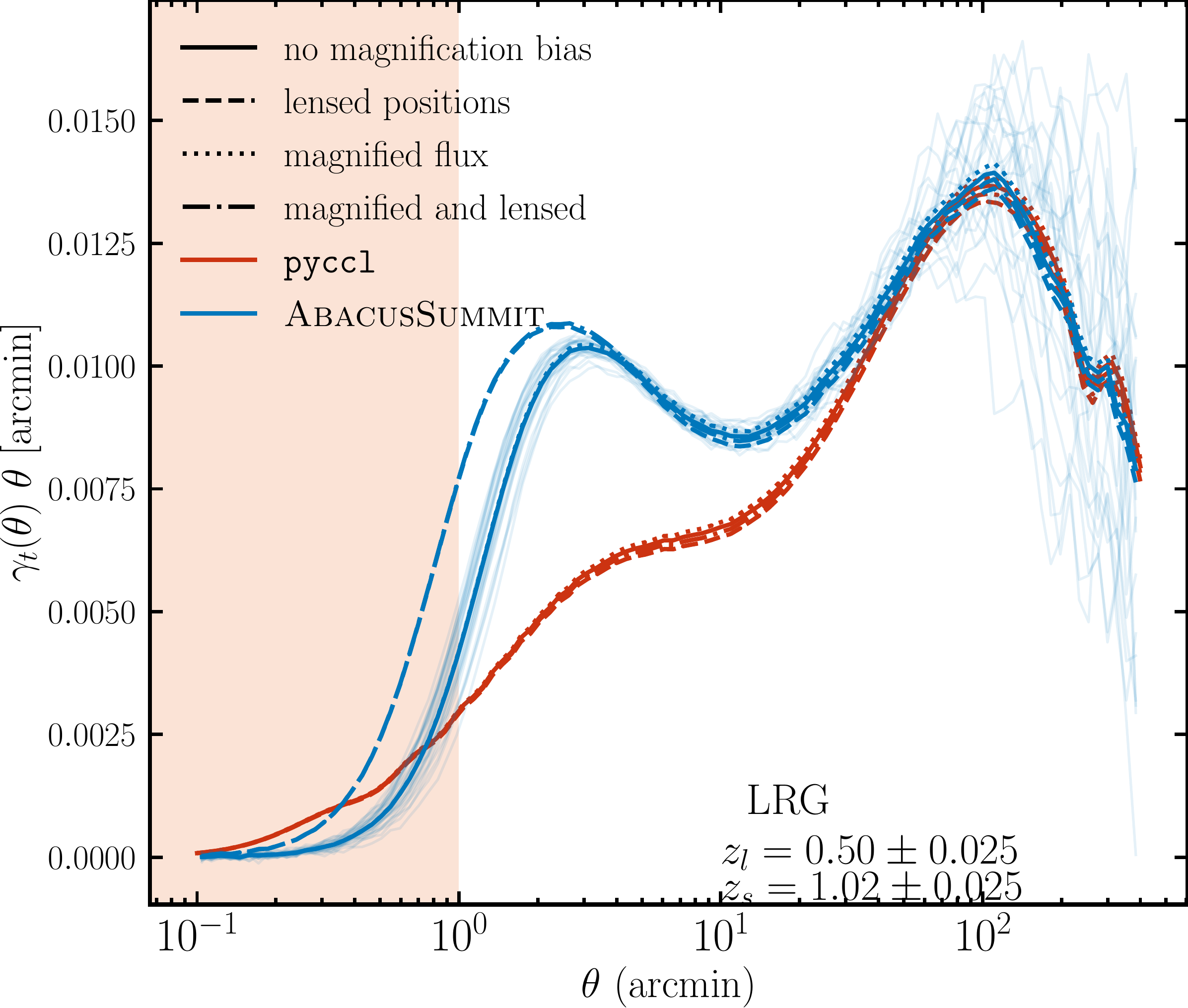}
    \includegraphics[width=0.44\textwidth]{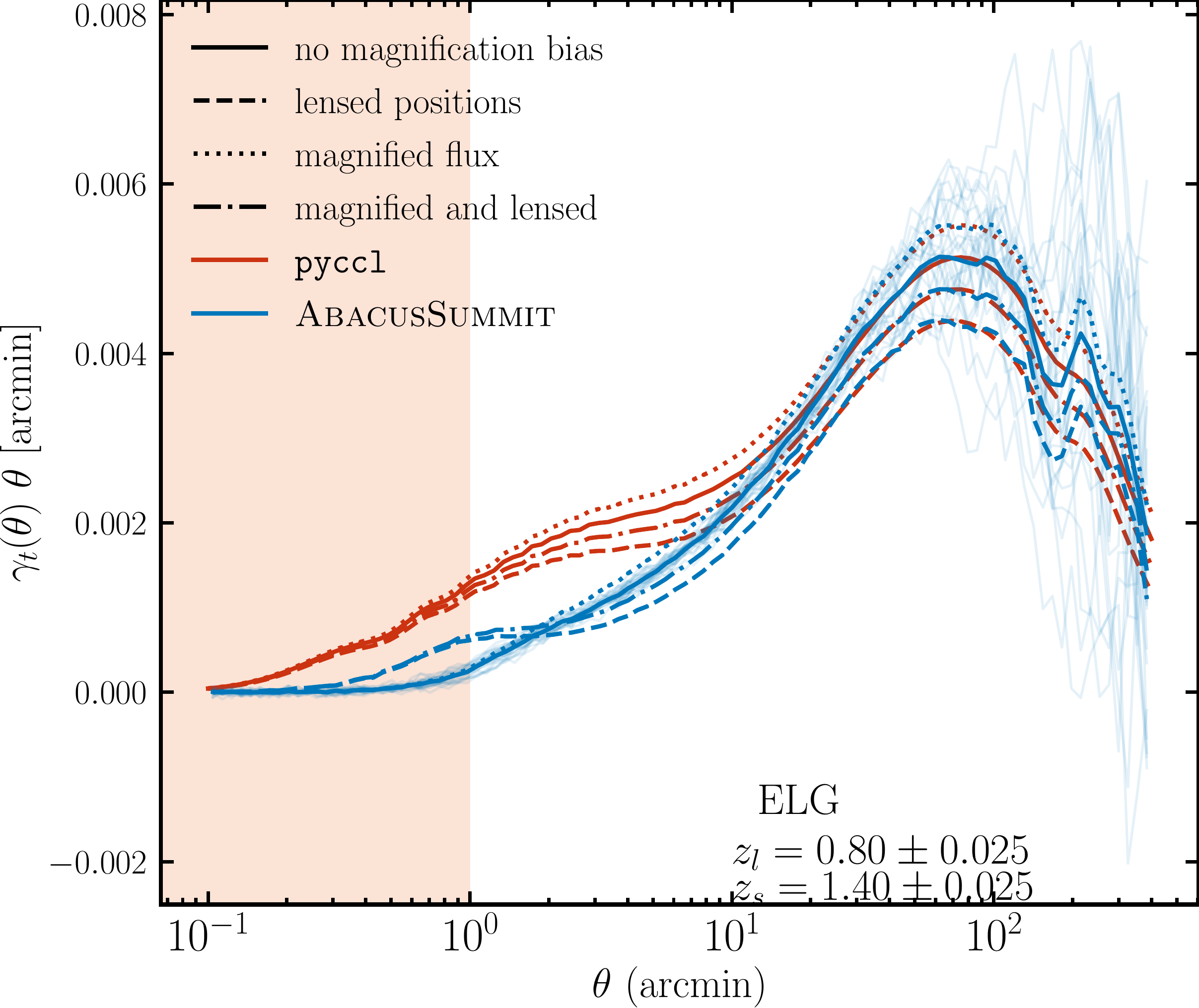}
    \includegraphics[width=0.44\textwidth]{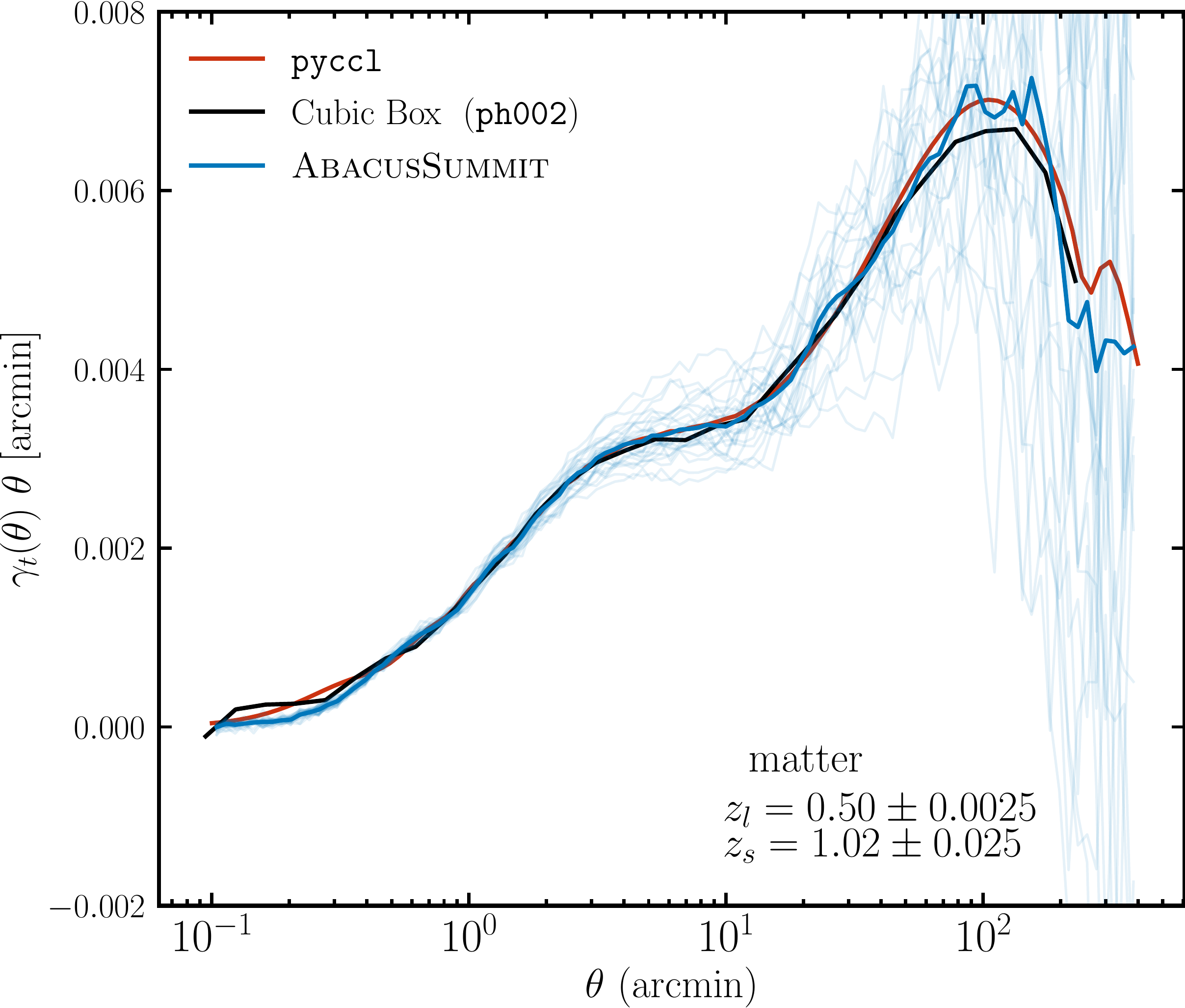}
    \caption{Galaxy-shear measurement for lens LRG (top) and ELG (middle) samples located at $z_l = 0.5$ and $z_s = 1$ or $z_l = 0.8$ and $z_s = 1.4$, in the case of no magnification bias (solid), lensed positions (dashed), magnified flux (dotted), and both (dash-dotted) (see Section~\ref{sec:implmag}). Magnification bias has the same response in the simulations (blue) as in theory (red). On large scales, the curves are matched well, but sample variance introduces substantial noise. On small scales, the one-halo term is overpredicted for ELGs and underpredicted for LRGs by the linear bias model. Results are shown for all 25 boxes. The red shaded regions corresponds (conservatively) to scales affected by the Born approximation. \boryana{On the bottom panel, we show the matter-shear measurement computed via our weak lensing mocks (blue), the cubic box (black) and \texttt{halofit} (red), demonstrating excellent agreement down to $r > 0.4 \ {\rm arcmin}$.}}
    \label{fig:gammat}
\end{figure}

When measuring galaxy-galaxy lensing in configuration space, it is convenient to work with the average tangential shear $\langle \gamma_t(\theta)\rangle $, which defines the tangential shear of background galaxies at angular separation $\theta$ from a lens galaxy. This quantity is related to the convergence $\kappa$ via $\langle \gamma_t (\theta) \rangle = \langle \overline{\kappa} (< \theta) - \langle \kappa (\theta) \rangle$ with $ \overline{\kappa} (< \theta) $ being the integrated convergence within separation $\theta$. Thus, we can express $\gamma_t$ in terms of the angular cross-power spectrum, $C_\ell^{\kappa g}$, as:
\begin{equation}
\label{eq:gammatamodel}
    \gamma_t (\theta) = \int \frac{d\ell \, \ell }{ 2 \pi} \, C_\ell^{\kappa g} \, J_2(\ell \theta) ,
\end{equation}
where $J_2$ denotes the second-order Bessel function of the first kind. 

In Fig.~\ref{fig:gammat}, we show the galaxy-shear cross-correlation measurement for the lens population of LRGs (ELGs) at $z_l = 0.5$ (0.8) and source population at $z_s = 1.025$ (1.4). We adopt the same redshift distribution for the lens population as in Fig.~\ref{fig:clkg}, i.e. Gaussian centered at $z = 0.5$ (0.8) with width of $\Delta_z = 0.2$. Additionally, we study the effect of magnification bias by switching on and off the effects of lensed positions and magnified flux, as described in Section~\ref{sec:implmag}. Qualitatively, the magnification bias curves behave as expected from theory: for our choice of $s = 0.2$ (see Section~\ref{sec:magbias}), the clustering increases due to flux magnification (effective change of depth) and decreases due to the lensing of the positions (effective change of area). On large scales, the curves are matched well, but sample variance introduces substantial noise at $\theta \gtrsim 100 \ {\rm arcmin}$. This echoes our findings in Fig.~\ref{fig:xipm}. Similarly to Fig.~\ref{fig:clkg}, we see that the one-halo term (and the one-to-two-halo transition) is overpredicted for ELGs and underpredicted for LRGs by the simple \texttt{halofit} model with linear bias. This is the result of the smaller mass (and thus, bias) and lower mean satellite occupation of ELG host haloes, which results in a lower one-halo clustering relative to the two-halo term (which is captured well by modulating the bias). The opposite is true for LRGs, which are hosted by haloes that are larger than the typical halo (and thus, more biased) and on average in the high-mass end, contain many more satellites. We note that the \texttt{pyccl} predictions use the same values of the bias as Fig.~\ref{fig:clkg}; namely, 1.1 ($z = 0.8$) and 1.25 ($z = 1.025$) for the ELGs, and 1.95 ($z = 0.5$) and 2.3 ($z = 1.025$) for the LRGs. We revisit the galaxy lensing measurement in Fig.~\ref{fig:gt_desy3}, where we explore the predicted signal for DES Y3 with realistic noise properties.

\boryana{In the bottom panel of Fig.~\ref{fig:gammat}, we show the particle- (matter-)shear measurement obtained from our weak lensing maps (averaging over all 25 simulations), the cubic box (by converting $\Delta_\Sigma(r_p)$ into $\gamma_t(\theta)$ for a single realization, \texttt{AbacusSummit\_base\_c000\_ph002}), and the \texttt{halofit} prediction from \texttt{pyccl}. The three curves show remarkable consistency with each other, which reassures us of the validity of our mocks. Below $r \lesssim 0.4$, we begin to see deviations between the weak lensing curve and the \texttt{halofit} prediction, which can be attributed to the breaking down of the Born approximation below scales associated with the root-mean-square of the deflection angle, $\alpha$. We note that in this calculation, we use the unlensed positions of both lenses and sources, whereas for the two galaxy panels above, we use the more `physically correct' lensed positions. However, we expect that in the single-plane (Born) approximation, the angular deflection by which we move the galaxies is not entirely accurate, leading to decorrelations between the matter field and the tracers, and thus a suppression in the measured clustering. The shaded band in the top two panels demarcates the region where we conservatively expect the Born approximation to break down ($r < 1 \ {\rm arcmin}$).}

In order to validate our lens population catalogues, we compute another useful projected galaxy clustering statistic, $w_p(r)$, defined as:
\begin{equation}
    w_p(r) = \sum_{\pi=0}^{\pi_{\rm max}}\left[\frac{DD(r, \pi)-2 DR(r, \pi)+RR(r, \pi)}{RR(r, \pi)} - 1\right],
\end{equation}
where $DD(r)$, $DR(r)$, and $RR(r)$ are the normalized data-data, data-random and random-random pair counts as a function of pair distance. For our test, we integrate pairs with line-of-sight separations out to $\pi_{\rm max} = 30\mpcoh$.

\begin{figure}
    \centering
    \includegraphics[width=0.48\textwidth]{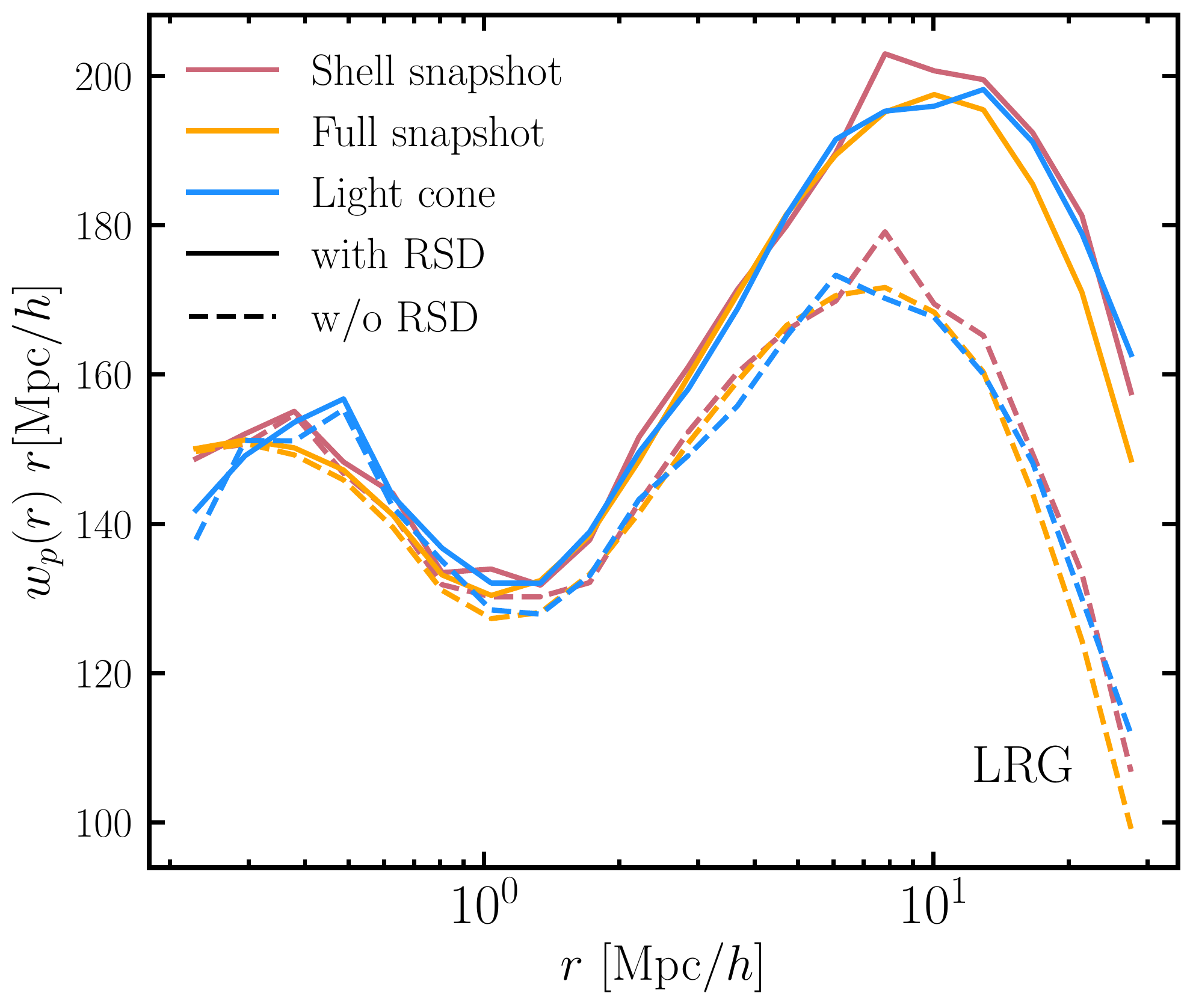}
    \includegraphics[width=0.48\textwidth]{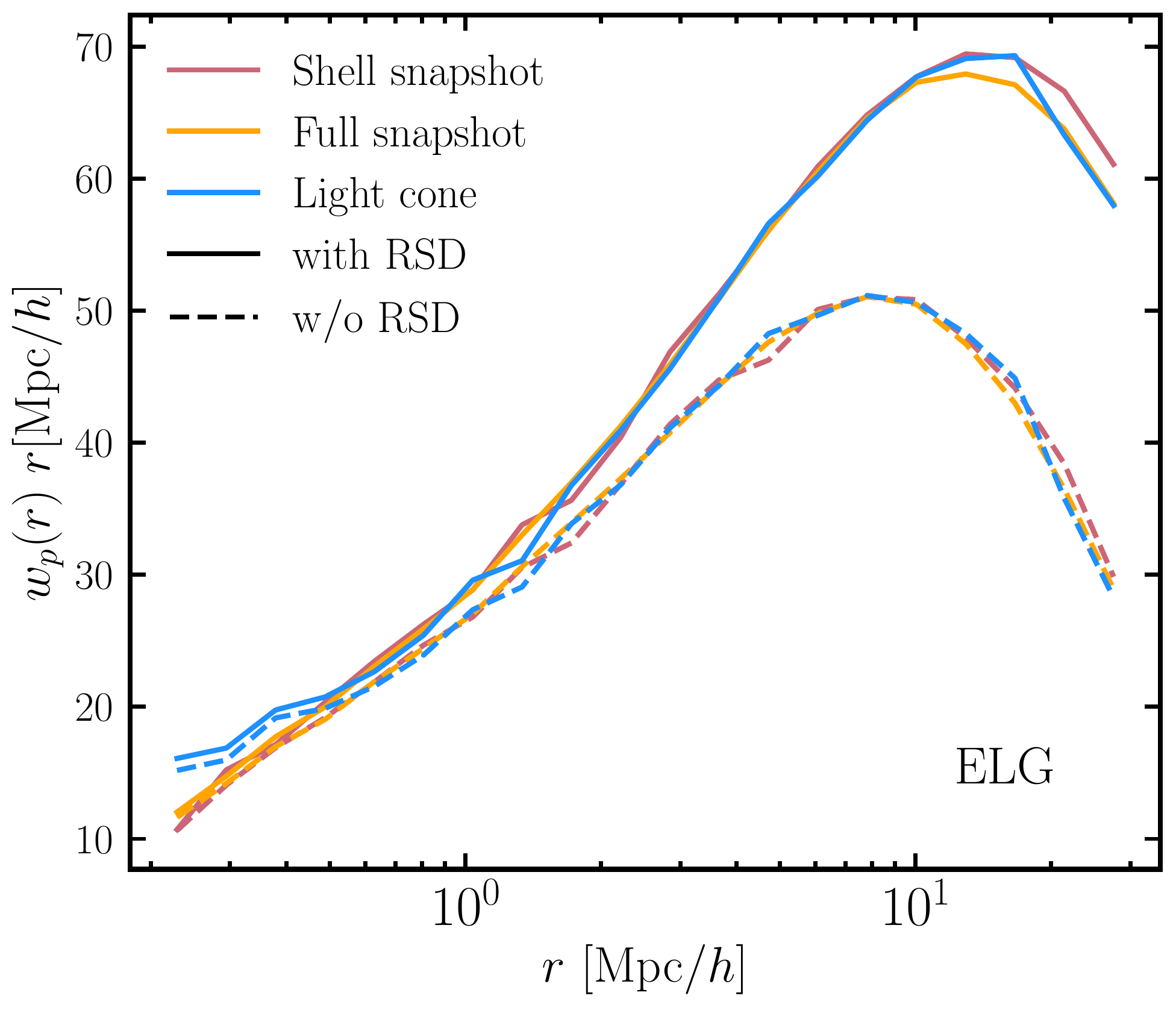}
    \caption{
    The galaxy projected auto-correlation function computed using the full periodic box (orange), the light cone galaxy catalogue (blue), and an octant shell of the periodic box (red) at the lens redshift, $z_l = 0.8$, integrated out to $\pi_{\rm max} = 30\mpcoh$. The top panel displays the LRG result, whereas the bottom one displays the ELG one (defined in Section~\ref{sec:galaxy}). We also show the effect of RSD on the clustering (solid and dashed lines). To isolate redshift evolution effects in the light cone catalogues, we select a thin slice around $z_l$ of thickness $\Delta z = 0.04$. We find that the agreement between all curves is excellent, validating our catalogue construction pipeline. Results are shown for \texttt{AbacusSummit\_base\_c000\_ph002}.}
    \label{fig:wp}
\end{figure}

In Fig.~\ref{fig:wp}, we show the projected auto-correlation function computed using the full periodic box, the light cone galaxy catalogue, and an octant shell of the periodic box (which has the same thickness and geometry as the light cone catalogue) at the lens redshift, $z_l = 0.8$. We choose this redshift, as it is the only redshift that the two tracers share in the GQC mocks \citep{Yuan+2023}. The top panel shows the result for the LRG galaxy catalogue, whereas the bottom one does so for the ELG one (see definitions of the two populations in Section~\ref{sec:galaxy}). We also test the effect of switching on and off RSD effects (solid and dashed lines, respectively) to make sure those are properly accounted for. To isolate redshift evolution effects in the light cone catalogues, we select a thin slice around $z_l$ of thickness $\Delta z = 0.04$. We adopt the corresponding comoving radial distance cut when defining the shell snapshot to ensure that the volume (and thus sample variance effects) are similar in both. We find that the agreement between all three samples for both galaxy populations is excellent, which validates both the light cone catalogue construction as well as the evolving HOD pipeline used to define the HOD parameters used in each light cone epoch (see Section~\ref{sec:galaxy} and Section~\ref{sec:hod}).

\section{Observations pipeline}
\label{sec:obs}

\subsection{Mock DESI samples}
\label{sec:redsh}

\begin{figure}
    \includegraphics[width=0.48\textwidth]{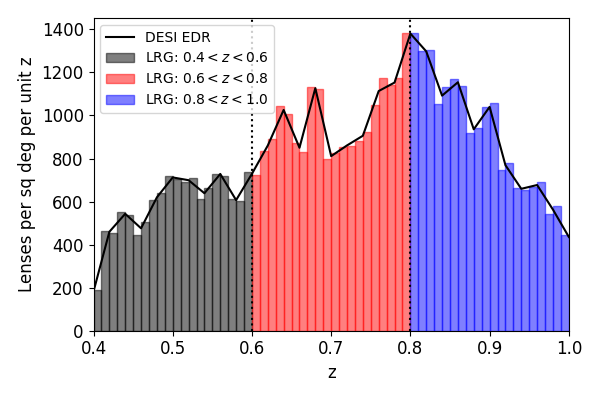}
    \caption{The angular density of selected DESI LRG galaxies in the \textsc{AbacusSummit} mock as a function of redshift, displayed as a histogram in $\Delta z = 0.01$ bins. Targets from the LRG samples are divided into three lens redshift samples $(0.4-0.6, 0.6-0.8, 0.8-1.0)$ as indicated by the histogram colours and vertical dashed lines.  The solid line indicates the measurement from the DESI Early Data Release.}
    \label{fig:nz_lens}
\end{figure}

We downsampled the galaxy catalogues produced via the procedures outlined in Section~\ref{sec:galaxy} to match the expected redshift distributions $N(z)$ of each respective survey.  In the case of DESI, we use the Early Data Release (EDR) catalogues to calculate the comoving number density as a \boryana{function of redshift \citep{Kremin+2023}}. The target $N(z)$ distribution of the DESI EDR is shown in Fig.~\ref{fig:nz_lens}, along with the redshift distribution of our downsampled samples. We produced catalogues in three redshift bins $0.4 < z < 0.6$, $0.6 < z < 0.8$ and $0.8 < z < 1.0$, which are designed to represent a future tomographic analysis of the DESI LRG sample.

We also created random catalogues matching the distribution of each mock DESI samples, for use in the correlation function measurements described in Sec.\ref{sec:cross}.  We drew the angular co-ordinates randomly from the light cone mask at reach redshift, and sampled random redshifts from the ensemble of data redshifts.

\subsection{Mock source samples}
\label{sec:real}

We assigned observational properties to our simulated lensing source populations, statistically drawn from the characteristics of the three weak lensing survey catalogues.  These properties include: photometric redshift, shape noise, shear calibration correction and source weight for weak lensing analyses.  In the following sub-sections we detail our implementations for each weak lensing survey, and validate that we successfully replicate the ensemble statistics of each survey.

For convenience, we created simulated shape catalogues with areas roughly matching the overlap of each weak lensing survey with DESI Y1 observations.  We implemented this partition by dividing the complete simulated source catalogues into angular pixels of area $\approx 53.7$ deg$^2$ using a \texttt{HEALPIX} pixelisation with resolution parameter $n_{\rm side} = 8$.  We then grouped these pixels into contiguous regions, using $(9, 16, 3)$ pixels to build each (KiDS, DES, HSC) mock shape catalogue.  We ensured that all contributing pixels possessed complete data along the light cone to $z=2$, such that each individual \textsc{AbacusSummit} mock can be used to produce $(4,2,12)$ catalogues representing the overlap of DESI Y1 and (KiDS, DES, HSC).

\subsubsection{KiDS}
\label{sec:kids}

We based our KiDS source mocks on the KiDS-1000 ``gold sample'' dataset \citep{2021A&A...645A.105G}, which consists of 21 million galaxies with an effective number density $n_{\rm eff} = 6.2$ gal/arcmin$^2$.  The shape catalogue is divided for analysis into five tomographic bins with edges $z_B = [0.1, 0.3, 0.5, 0.7, 0.9, 1.2]$, based on point photometric redshift estimates $z_B$ of the \texttt{BPZ} method.  Ellipticity measurements and weights for each source were provided by a self-calibrating version of \texttt{lensfit} \citep{2017MNRAS.467.1627F}.

We first sub-sampled the \textsc{AbacusSummit} source halo catalogues in a series of narrow redshift slices, using a minimum halo mass to match the source density of the KiDS-1000 sample in those slices calibrated by \cite{2021A&A...647A.124H}.  In order to re-create the division into tomographic source bins, we then statistically assigned photometric redshifts $z_p$ to each mock source from a scattering probability distribution $P(z_p|z_s)$ conditional on the ``spectroscopic'' or true simulation redshift $z_s$.  We built this probability map using the redshift calibration reference sample built by the KiDS collaboration \citep{2021A&A...647A.124H}, which is re-weighted using a self-organising map to closely resemble the distribution of KiDS-1000 sources.\footnote{We are grateful to Hendrik Hildebrandt for sharing this catalogue for the purpose of constructing these mocks.}

Next, we distorted the noise-free shear components from the simulated lensing fields to account for multiplicative shear calibration factors present in the shape catalogues, and statistical shape noise error.  Following Table 1 of \cite{2021A&A...645A.105G}, we assigned shear calibration corrections $m = [-0.009, -0.011, -0.015, 0.002, 0.007]$ and ellipticity dispersions per component $\sigma_e = [0.270, 0.258, 0.273, 0.254, 0.270]$ in the five tomographic bins.  We first applied the shear calibration correction by multiplying the noise-free shear components in each bin by $(1+m)$.  We then introduced shape noise to the source catalogues by determining the complex noisy shear $e = (\gamma+n)/(1+n \, \gamma^*)$ \citep{1997A&A...318..687S}, where the components of observed shear $(e_1,e_2)$ are found as $e = e_1 + i e_2$, the noise-free shear $\gamma = \gamma_1 + i \gamma_2$, and the noise $n = n_1 + i n_2$. The noise components $(n_1, n_2)$ are drawn from Gaussian distributions with standard deviation $\sigma_e$.  Finally, we assigned weights to each lensing source by randomly sub-sampling from the \texttt{lensfit} weights of the survey dataset within each tomographic bin.

This procedure resulted in a suite of KiDS source mocks with ensemble properties representing those of the real catalogues in each tomographic bin: redshift distribution, source weights, multiplicative shear calibration factors and shape noise.  We repeated this procedure to create DES and HSC mocks, emphasising some aspects specific to these surveys in the following sub-sections.

\subsubsection{DES}
\label{sec:des}

We based our DES source mocks on the DES Year 3 (DES-Y3) weak lensing shape catalogue \citep{2021MNRAS.504.4312G}, which was derived using the \textsc{metacalibration} pipeline, and consists of $100.2$ million galaxies with an effective number density $n_{\rm eff} = 5.6$ gal/arcmin$^2$.  Galaxies in the DES-Y3 dataset are clustered by a self-organising map into different ``phenotypes'', which are used to construct four tomographic source bins with edges $z_p = [ 0.0, 0.36, 0.63, 0.87, 2.0 ]$ \citep{2021MNRAS.505.4249M}.  In the \textsc{metacalibration} framework, all sources are assigned a shear response matrix $\mathbf{R}$ \citep{2021MNRAS.504.4312G}, which we approximate for our purposes by the leading-order diagonal elements $R_{11}$ and $R_{22}$ such that the shear calibration correction is a scalar quantity $\left( R_{11} + R_{22} \right)/2$.  The DES-Y3 framework also allows for a residual multiplicative bias with best values $m = [-0.006, -0.020, -0.024, -0.037]$ in the four tomographic bins \citep{2022PhRvD.105b3514A}.

After sub-sampling the mock halo catalogues as a function of true redshift $z_s$ to match the DES-Y3 source redshift distribution, we statistically assigned a tomographic bin $b = (1,2,3,4)$ to each mock source using a scattering probability distribution $P(b|z_s)$ constructed from the source density and calibrated redshift distribution in each tomographic bin.  We randomly assigned values of $R_{11}$, $R_{22}$ and weight to each mock source, drawn from the real DES-Y3 data catalogue in each tomographic bin, and applied the shear calibration correction by multiplying the noise-free shear components in each bin by $(1+m) \times \left( R_{11} + R_{22} \right)/2$.  Finally, we applied shape noise to the distorted noise-free shear components using the formulation described in Sec.~\ref{sec:kids}, where we take $\sigma_e = [0.201, 0.204, 0.195, 0.203]$ in the four tomographic bins.  These noise values differ from those quoted in Table 1 of \citep{2022PhRvD.105b3514A}, because we are applying them to ellipticities which have already been distorted by the shear responsivity.  The overall shape noise, including calibration factors, matches the noise properties of DES-Y3.

This procedure resulted in a suite of DES-Y3 source mocks with assigned tomographic bins, distorted and noisy ellipticities, shear calibration factors $(R_{11}, R_{22})$, and lensing weights.

\subsubsection{HSC}
\label{sec:hsc}

We based our HSC mocks on the HSC-Y1 weak lensing shape catalogue \citep{2018PASJ...70S..25M}, which comprises 9 million galaxies with an effective number density $n_{\rm eff} = 17.6$ gal/arcmin$^2$.  Sources are assigned into four tomographic bins using point estimates from the \texttt{Ephor} photo-$z$ code \citep{2019PASJ...71...43H}.  The shear estimators in HSC use a responsivity factor $2R$ (see Eq.3 in \cite{2018PASJ...70S..25M}), where $R = 1 - e^2_{\rm rms}$ and $e_{\rm rms}$ is the RMS intrinsic distortion per component, along with a residual multiplicative bias $m$.  Finally, sources are assigned an inverse-variance shape weight $w$ based on a quadrature sum of shape noise and measurement error \citep{2018PASJ...70S..25M}.

As described in Sec.~\ref{sec:kids}, we statistically assigned photometric redshifts to each mock source from a scattering probability distribution calibrated by the HSC redshift calibration sample \citep{2019PASJ...71...43H}.  We randomly assigned values of shape weight $w$, $e_{\rm rms}$ and $m$ to each mock source, drawn from the real HSC data catalogue in each tomographic bin, and applied the shear calibration correction by multiplying the noise-free shear components in each bin by $(1+m) \times 2 \left( 1 - e^2_{\rm rms} \right)$.  Finally, we applied the noise to the distorted noise-free shear components, where we deduced the noise to be applied to each source as $\sigma_e = 1/\sqrt{w}$ (which combines both shape noise and measurement error).

This procedure resulted in a suite of HSC source mocks with assigned tomographic bins, distorted and noisy ellipticities, shear calibration factors $(e_{\rm rms}, m)$, and lensing weights.

\subsubsection{Summary statistics}
\label{sec:lensingstats}

\begin{figure*}
    \includegraphics[width=0.9\textwidth]{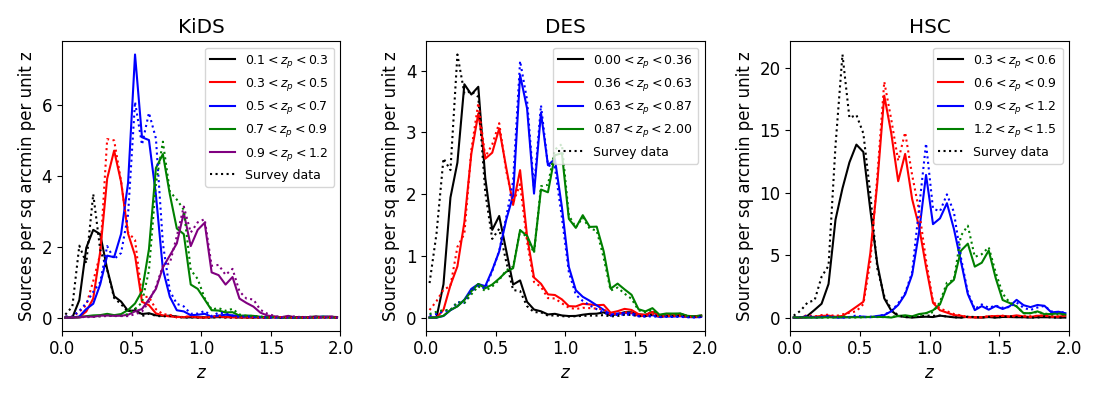}
    \caption{The angular density of mock source galaxies in the KiDS (left panel), DES (middle panel) and HSC (right panel) \textsc{AbacusSummit} mocks as a function of spectroscopic redshift.  Separate redshift distributions are plotted for each source tomographic sample, and the measurements from the mocks (solid lines) and compared to the redshift probability distributions defined for the data samples by each survey collaboration (dotted lines).}
    \label{fig:nz_source}
\end{figure*}

Before testing the correlation properties of these source mocks, we first validated that their ensemble statistical properties in each tomographic bin matched each weak lensing survey.  Fig.~\ref{fig:nz_source} displays the spectroscopic redshift distributions of each separate tomographic sample of the KiDS, DES and HSC source mocks.  The distributions are compared to those defined for the data samples by each survey collaboration, demonstrating that the source mocks are representative of the data.  The only significant discrepancy is found at the lowest redshifts, where there is a small deficit of mock sources because the mass resolution of the mocks is insufficient to match the number density of the deepest weak lensing catalogues at low redshift.  However, low-redshift sources have little constraining power for weak lensing, such that this deficit is unimportant.  The ``spikes'' in the redshift distribution of Fig.~\ref{fig:nz_source} correspond to sample variance fluctuations in the spectroscopic reference sets used by each lensing collaboration.

\begin{figure*}
    \includegraphics[width=0.7\textwidth]{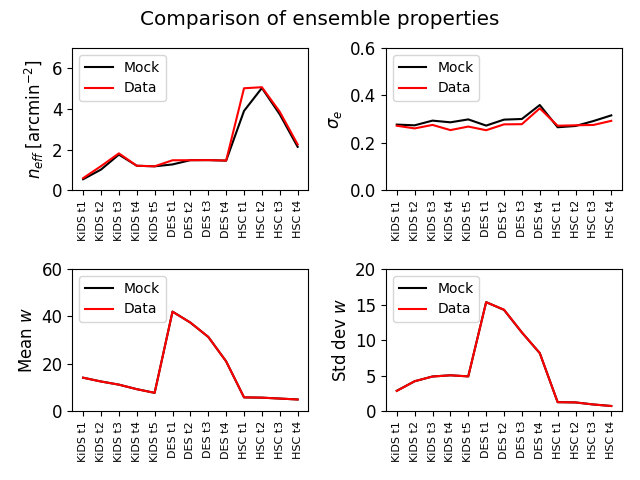}
    \caption{Comparison of some ensemble properties of the KiDS, DES and HSC \textsc{AbacusSummit} source mocks, and the corresponding data catalogues of those surveys.  Results are displayed for each source tomographic sample (5 for KiDS, 4 for DES and 4 for HSC).  The ensemble properties considered are the effective source number density $n_{\rm eff}$ (top-left panel), the effective shape noise $\sigma_e$ (top-right panel), the mean source weight (bottom-left panel), and the standard deviation of the source weight across the sample (bottom-right panel).}
    \label{fig:mock_vs_data_comparison}
\end{figure*}

Fig.~\ref{fig:mock_vs_data_comparison} compares some ensemble properties of each separate tomographic sample of the KiDS, DES and HSC \textsc{AbacusSummit} source mocks, and the corresponding survey data catalogues.  The ensemble properties considered are the effective source number density, effective shape noise, average source weight and standard deviation of the source weight.  The effective number density is defined in terms of the source weights $w_i$ for each galaxy, following \cite{2012MNRAS.427..146H}, by
\begin{equation}
n_{\rm eff} = \frac{1}{A} \frac{\left( \sum_i w_i \right)^2}{\sum w_i^2} ,
\end{equation}
where $A$ is the area covered by the catalogue, and the effective shape variance by,
\begin{equation}
\sigma^2_e = \frac{1}{2} \left[ \frac{\sum_i w^2_i e^2_{i,1}}{\sum_i w^2_i} + \frac{\sum_i w^2_i e^2_{i,2}}{\sum_i w^2_i} \right] .
\end{equation}
Fig.~\ref{fig:mock_vs_data_comparison} shows good agreement between the properties of the mock and data catalogues.

\subsection{Correlation function tests of the mock samples}
\label{sec:cross}

We validated the \textsc{AbacusSummit} weak lensing and DESI mock catalogues by comparing their measured 2-point correlation functions to theoretical predictions.  We note that full cosmological parameter fits with analytical covariance are beyond the scope of the current study, and will be presented by future work.  The scope of this section is to simply present the correlation function measurements, and check for any significant discrepancies with theoretical predictions that would indicate serious failures in mock construction.

\begin{figure*}
    \includegraphics[width=0.9\textwidth]{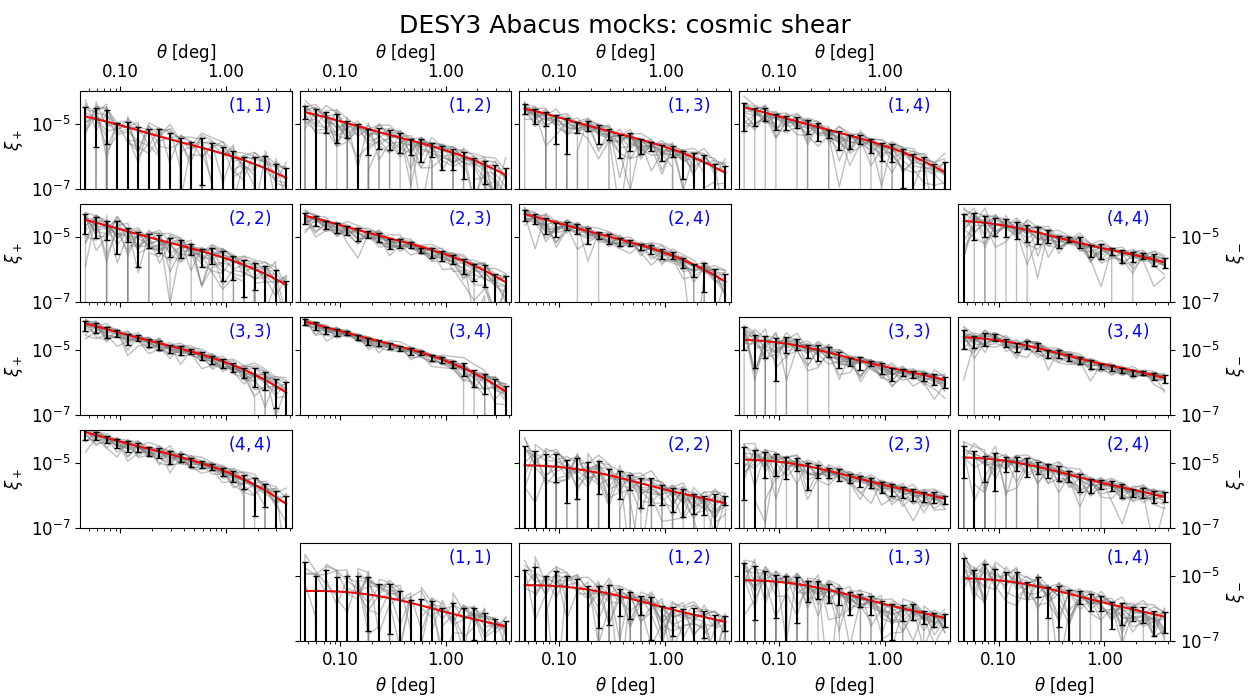}
    \caption{The shear correlation functions $(\xi_+, \xi_-)$ between the four tomographic source samples of the DES \textsc{AbacusSummit} mocks, as a function of angular separation $\theta$ in degrees.  The different panels display measurements for different combinations of tomographic samples, as indicated by the identifiers in the top-right corner of each panel.  We plot the mean and standard deviation of the measurements across the \textsc{AbacusSummit} DES regions, where the light grey lines indicate measurements for a selection of individual regions. The solid lines show the fiducial cosmological models for $(\xi_+, \xi_-)$.}
    \label{fig:xipm_desy3}
\end{figure*}

First, we measured the cosmic shear correlation functions $\xi_\pm(\theta)$ of the different source catalogues in tomographic bins, using the \textsc{TreeCorr} package \citep{2015ascl.soft08007J}.  We performed measurements using the same angular separation binning adopted by each weak lensing collaboration when presenting their cosmic shear results.  Fig.~\ref{fig:xipm_desy3} displays the shear correlation function measurements for the DES mocks in four tomographic bins (results for the KiDS and HSC mocks have a similar presentation, and we do not show them).  We plot the mean and standard deviation of the measurements for the different DES regions (which have areas representative of the overlap of the DES Y3 and DESI Y1 datasets), compared to the fiducial cosmological models in each case, computed from the source redshift distributions and \textsc{AbacusSummit} matter power spectra.  The measured correlation functions are corrected for the applied shear calibration factors included in the mock outputs.  We find that the fiducial models provide a good description of these measurements.

\begin{figure*}
    \includegraphics[width=0.9\textwidth]{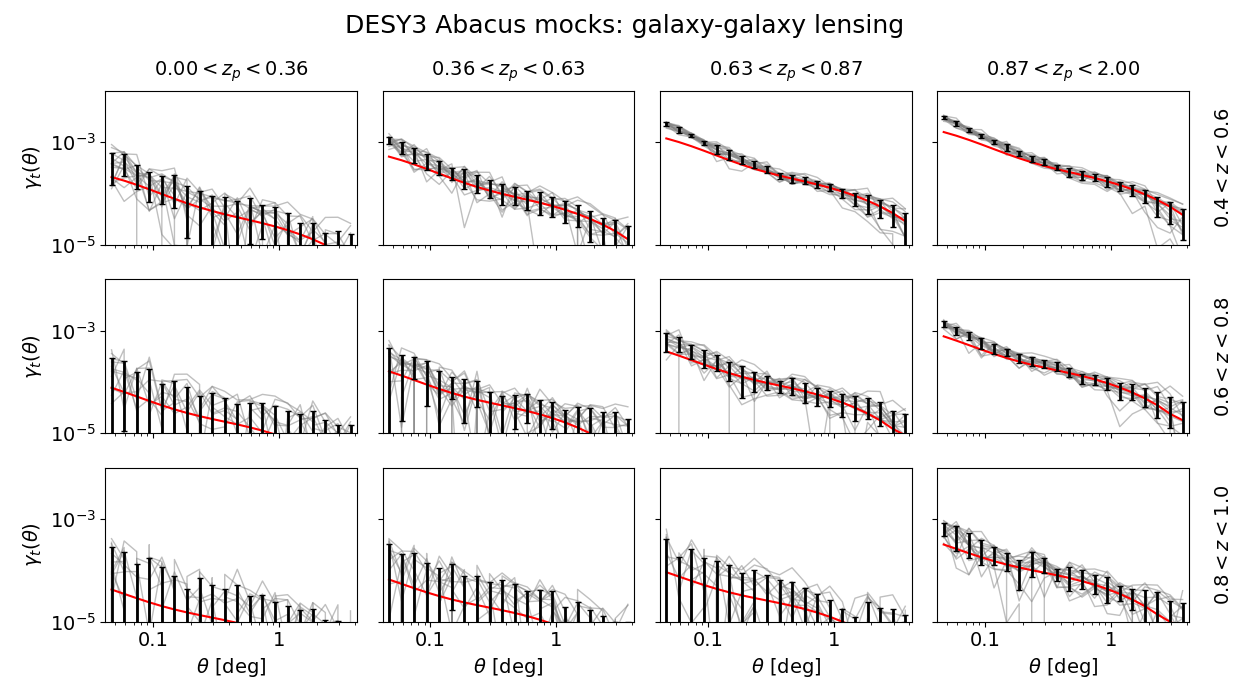}
    \caption{The average tangential shear $\gamma_t$ of the four tomographic source samples of the DES \textsc{AbacusSummit} mocks, around the three lens samples of the DESI \textsc{AbacusSummit} mocks, as a function of angular separation $\theta$ in degrees.  The different panels display measurements for different combinations of source and lens samples, as indicated by the captions above and to the right of each panel.  We plot the mean and standard deviation of the measurements across the DES regions, where the light grey lines indicate measurements for a selection of individual regions.  The solid lines show the fiducial cosmological models for $\gamma_t$, assuming a representative linear galaxy bias factor.}
    \label{fig:gt_desy3}
\end{figure*}

Next, we measure the average tangential shear (galaxy-galaxy lensing signal) $\gamma_t(\theta)$ between the mock source catalogues and the lens samples of the DESI \textsc{AbacusSummit} mocks, again using \textsc{TreeCorr} and adopting the same angular separation bins as before.  Fig.~\ref{fig:gt_desy3} displays these measurements between the four tomographic source samples of the DES mocks, around the three lens samples of the DESI mocks (results for the KiDS and HSC mocks again have a similar appearance).  We plot the mean and standard deviation of the measurements for the different DES regions, compared to the fiducial cosmological models in each case, assuming a representative linear bias factor $b=2$.  We again note that the fiducial models provide a good description of these measurements, except at small scales, where the linear bias formulation is inaccurate.

\begin{figure*}
    \includegraphics[width=0.9\textwidth]{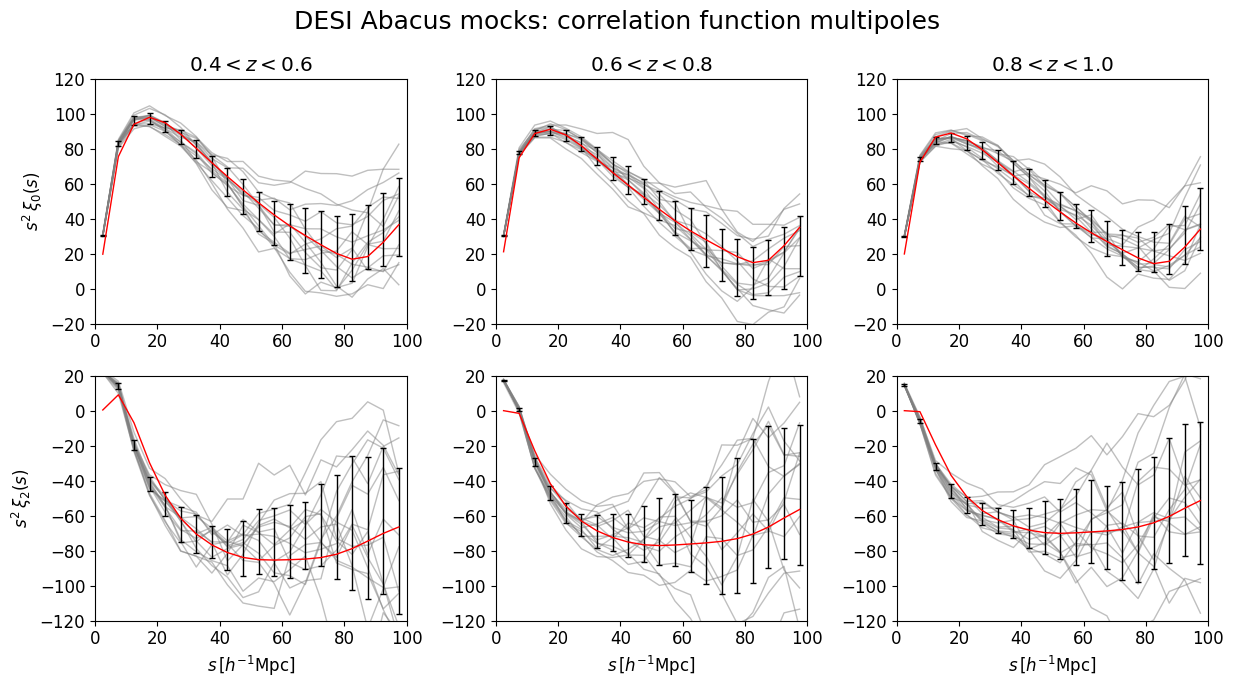}
    \caption{The monopole and quadrupole correlation function $\xi_0$ and $\xi_2$ of the three lens samples of the DESI \textsc{AbacusSummit} mocks, as a function of spatial separation $s$ in units of $\mpcoh$.  We plot the mean and standard deviation of the measurements across the DES regions, where the light grey lines indicate measurements for a selection of individual regions.  The solid lines show the fiducial cosmological models, assuming a representative linear galaxy bias factor and RSD streaming model.}
    \label{fig:xipole_desy3}
\end{figure*}

Finally, we measure the correlation function multipoles of the DESI \textsc{AbacusSummit} mocks in the three redshift bins $(0.4-0.6, 0.6-0.8, 0.8-1.0)$, using the \textsc{CorrFunc} package \citep{2020MNRAS.491.3022S}.  We perform monopole and quadrupole measurements in 20 linearly-spaced separation bins between 0 and $100\mpcoh$.  Fig.~\ref{fig:xipole_desy3} displays these correlation function multipole measurements for the DESI mocks in the DES regions.  We plot the mean and standard deviation of the measurements for the different regions, compared to the fiducial cosmological models in each case, assuming a representative linear bias factor $b=2$ and RSD streaming model with small-scale velocity dispersion $\sigma_v = 300$ km s$^{-1}$.  The fiducial models provide a good description of the signals, except at small scales where the linear RSD prescription breaks down.

\section{Summary}
\label{sec:conc}

In this paper, we present high-resolution curved-sky weak lensing maps, generated using the \boryana{Born approximation} and the $N$-body simulation suite \textsc{AbacusSummit}, as well as accompanying weak lensing mock catalogues, which are tuned via fits to DESI small-scale clustering measurements of LRGs and ELGs. We make these products publicly available under \href{https://app.globus.org/file-manager?origin_id=fc0006ee-68fc-11ed-8fd2-e9cb7c15c7d2&origin_path=%2F}{this URL}. The purpose of these products is to aid the joint analysis between galaxy surveys such as DESI and weak lensing surveys such as HSC, DES, and KiDS.

In Section~\ref{sec:lens}, we provide a detailed description of our procedure for generating the weak lensing maps, which follows the ``Onion Universe'' approach \citep{2015MNRAS.447.1319F}, as well as various validation plots. The available maps consist of the cosmic shear, deflection angle and convergence fields at source redshifts ranging from $z = 0.15$ to 2.45 with $\Delta z = 0.05$ as well as CMB convergence maps ($z \approx 1089.3$) for each of the 25 \texttt{base}-resolution simulations ($L_{\rm box} = 2000\mpcoh$, $N_{\rm part} = 6912^3$) as well as for the two \texttt{huge} simulations ($L_{\rm box} = 7500\mpcoh$, $N_{\rm part} = 8640^3$) at the fiducial AbacusSummit cosmology ($Planck$ 2018). The pixel resolution of each map is 0.21 arcmin, corresponding to a HEALPiX $N_{\rm side}$ of 16384. The sky coverage of the \texttt{base} simulations is an octant until $z \approx 0.8$ (decreasing to about 1800 deg$^2$ at $z \approx 2.4$), whereas the \texttt{huge} simulations offer full-sky coverage until $z \approx 2.2$. 

As validation of the source maps and catalogues, we study the angular power spectrum of the convergence field, $C_\ell^{\kappa \kappa}$, shown in Fig.~\ref{fig:clkk}, and the shear auto-correlation, $\xi_\pm (\theta)$, shown in Fig.~\ref{fig:xipm}. We find that the base-resolution simulations show a remarkable agreement with the theoretical prediction for $C_\ell^{\kappa \kappa}$ from \texttt{pyccl}, which in turn employs the \texttt{halofit} matter power spectrum. The \texttt{huge}-resolution result displays a larger deviation from \texttt{pyccl} (though is in great agreement with Cosmic Emu), which we attribute to resolution effects and explore further in Appendix~\ref{app:halofit}. The shear-shear angular correlation function from \textsc{AbacusSummit} also shows excellent agreement with theory in most regimes, but suffers from boundary effects in the $\xi_+ (\theta)$ measurement for $\theta > 100$ arcmin. We note, however, that the errors are still within the noise of the weak lensing surveys of interest to this work, and the deficit is not visible when mimicking these surveys near $\theta \sim 200$ arcmin (see Fig.~\ref{fig:xipm_desy3}). These are investigated in detail in Appendix~\ref{app:shear}. We comment that the cutoff of $\xi_-$ relative to \texttt{pyccl} at $\theta < 1$ arcmin is \boryana{due to resolution effects (i.e., the pixel size of 0.21 arcmin), which are pushed to larger scales for $\xi_-$ than $\xi_+$.}

In Section~\ref{sec:mock}, we describe our process for assigning lensing properties ($\gamma_{1, 2}$) to our source population, which covers the redshift range between $z = 0.15$ and 2, as well as our implementation of magnification bias for the lens population of LRGs and ELGs, which covers the redshift range between $z = 0.3$ and 1.4. To enable the application of magnification bias to our HOD catalogues, we record the lensed positions and appropriate weights accounting for flux magnification for each lens galaxy. We then validate the mock creation procedure by comparing to \texttt{pyccl}, which adopts a simple linear bias approximation. As expected, we find reasonable agreement on large scales between the galaxy-galaxy lensing observables, $C_\ell^{\kappa g}$ and $\gamma_t (\theta)$, but on small scales, the linear bias model shows substantial deviations, overpredicting the clustering of ELGs and underpredicting that of the LRGs (see Fig.~\ref{fig:clkg} and Fig.~\ref{fig:gammat}). Furthermore, we generate random catalogues for the lens population of galaxies, which can be used to measure the galaxy auto-correlation. In Fig.~\ref{fig:wp}, we show the projected galaxy clustering, $w_p(r)$, with and without redshift space distortions for both LRGs and ELGs, and compare it with two samples generated using the periodic box, finding exquisite agreement.

In Section~\ref{sec:obs}, we sub-sample the \textsc{AbacusSummit} halo mocks to match the redshift distributions of the DESI Early Data Release, and the ensemble properties of the KiDS, DES and HSC weak lensing datasets.  We perform a statistical assignment of photometric redshifts to partition the data into tomographic bins matching those chosen by the lensing survey collaborations, and we also apply shear calibration corrections and shape noise in the framework of these surveys.  Finally, we verify that the cosmic shear correlation functions, galaxy-galaxy lensing signal and clustering multipole correlation functions of these catalogues agree with those predicted in the fiducial cosmology.

An immediate application of our products, planned for the near-term, is to extensively test the pipelines developed for analyzing redshift and weak lensing surveys, and ensure they are capable of correctly recovering the underlying cosmology and robust to various systematic and observational effects. A longer-term goal is to construct an emulator and use it to constrain astrophysical (e.g., galaxy-halo connection) and cosmological parameters from combinations of the latest galaxy and weak lensing data sets. \boryana{We are also planning to implement beyond-Born approximation corrections and baryonic effects, which would allow us to more reliably model $\sim$0.1 arcmin scales.}

\section*{Acknowledgements}

We would like to express our thanks to Martin Reinecke, Joe DeRose, Benjamin Joachimi, and David Alonso, for illuminating discussions during the preparation of this manuscript.  We thank Hendrik Hildebrandt for sharing the KiDS redshift calibration catalogue for the purpose of mock construction.

This research is supported by the Director, Office of Science, Office of High Energy Physics of the U.S. Department of Energy under Contract No. DE–AC02–05CH11231, and by the National Energy Research Scientific Computing Center, a DOE Office of Science User Facility under the same contract; additional support for DESI is provided by the U.S. National Science Foundation, Division of Astronomical Sciences under Contract No. AST-0950945 to the NSF’s National Optical-Infrared Astronomy Research Laboratory; the Science and Technologies Facilities Council of the United Kingdom; the Gordon and Betty Moore Foundation; the Heising-Simons Foundation; the French Alternative Energies and Atomic Energy Commission (CEA); the National Council of Science and Technology of Mexico (CONACYT); the Ministry of Science and Innovation of Spain (MICINN), and by the DESI Member Institutions: \url{https://www.desi.lbl.gov/collaborating-institutions}.

The authors are honored to be permitted to conduct scientific research on Iolkam Du’ag (Kitt Peak), a mountain with particular significance to the Tohono O’odham Nation.

\section*{Data Availability}

The data and accompanying scripts are publicly available and stored as a NERSC shared endpoint, ``AbacusSummit Weak Lensing'', at \url{https://app.globus.org/file-manager?origin_id=3dd6566c-eed2-11ed-ba43-09d6a6f08166&path=\%2F}. Data points for the figures are available at \url{https://doi.org/10.5281/zenodo.7926269}.



\bibliographystyle{mnras}
\bibliography{example} 




\appendix



\section{Testing the partial sky shear maps}
\label{app:shear}

Given the unusual geometric configuration of the \textsc{AbacusSummit} \texttt{base} light cones (see Fig.~\ref{fig:geo}), it is important to test the boundary effects on the lensing observables derived through harmonic transforms of the convergence field, i.e. the shear and deflection angle field. To this end, we construct the following simple test using the \texttt{huge} simulations, for which we know the ``truth,'' since they cover the entire sky rather than an octant. 

\textbf{Test 1}

First, we test the effect of the $\kappa(\hat n)$ to $\gamma_{1,2} (\hat n)$ conversion in the event of incomplete sky coverage. In particular, we measure the $\xi_\pm (\theta)$ signal for four different scenarios. 
\begin{enumerate}
    \item Scenario 1: measuring $\xi_\pm$ from the full-sky $\gamma_{1,2} (\hat n)$ map, which has been obtained from the full-sky $\kappa (\hat n)$ following Eq.~\ref{eq:shear}.
    \item Scenario 2: measuring $\xi_\pm$ from only an octant of the full-sky $\gamma_{1,2} (\hat n)$ map, obtained from the full-sky $\kappa (\hat n)$.
    \item Scenario 3: measuring $\xi_\pm$ from the octant $\gamma_{1,2} (\hat n)$ map, obtained by converting an octant of $\kappa (\hat n)$ into $\gamma_{1,2} (\hat n)$ following Eq.~\ref{eq:shear}.
	\item Scenario 4: measuring $\xi_\pm$ from the octant $\gamma_{1,2} (\hat n)$ map, obtained from an octant of $\kappa (\hat n)$ multiplied by a mask smoothed at the boundaries with a Gaussian kernel of 2 deg.
\end{enumerate}

In Fig.~\ref{fig:xipm_v1}, we show the results of this exercise, finding that the dominant effect causing deviations from theory is cosmic variance, though the partial sky conversion does affect the largest scales. In particular, the curves exhibit minimal differences for $\theta < 100$ arcmin for $\xi_\pm (\theta)$, and we only see noticeable deviations past that scale for $\xi_+ (\theta)$. As expected, the full-sky gamma measurements of $\xi_\pm (\theta)$ agree best with the theoretical prediction, as they do not suffer from boundary effects. We further notice that the maps obtained by converting an octant of the $\kappa(\hat n)$ field into $\gamma_{1,2} (\hat n)$ yield almost identical results regardless of whether a smoothed mask is applied or not. Finally, the measurements obtained from an octant $\gamma_{1,2} (\hat n)$ map cutout from the full-sky shear maps exhibit only marginal improvement relative to the other two octant results, suggesting that the main contributor to the noise is sample variance at those scales. We investigate this further in \textbf{Test 2}. We comment that on the smallest scales, $\theta < 1$ arcmin, the $\xi_- (\theta)$ measurements appear to be noise-dominated. We conjecture that this feature is due to the use of random halo particles as sources, the nearest pixel assignment procedure, and the map resolution of 0.2 arcmin.

\textbf{Test 2}

The second test we perform checks separately the octant conversion procedure by using all 8 octants. We compute two sets of 8 measurements, one for each octant. In the first set, we split the full-sky shear maps, $\gamma_{1,2} (\hat n)$, into 8 $\gamma_{1,2} (\hat n)$ submaps covering an octant of the sky. This provides us the ``idealized'' test case, in which the shear maps have been obtained from the full-sky convergence maps. In the second set, we apply the same methodology as we do for the base-resolution simulations: namely, after splitting the convergence map into octants, we apply the appropriate $\ell$-filter (see Eq.~\ref{eq:shear}) to each submap to obtain the shear. We then calculate $\xi_\pm (\theta)$ for the octants in both sets. 

\begin{figure}
    \centering
    \includegraphics[width=0.48\textwidth]{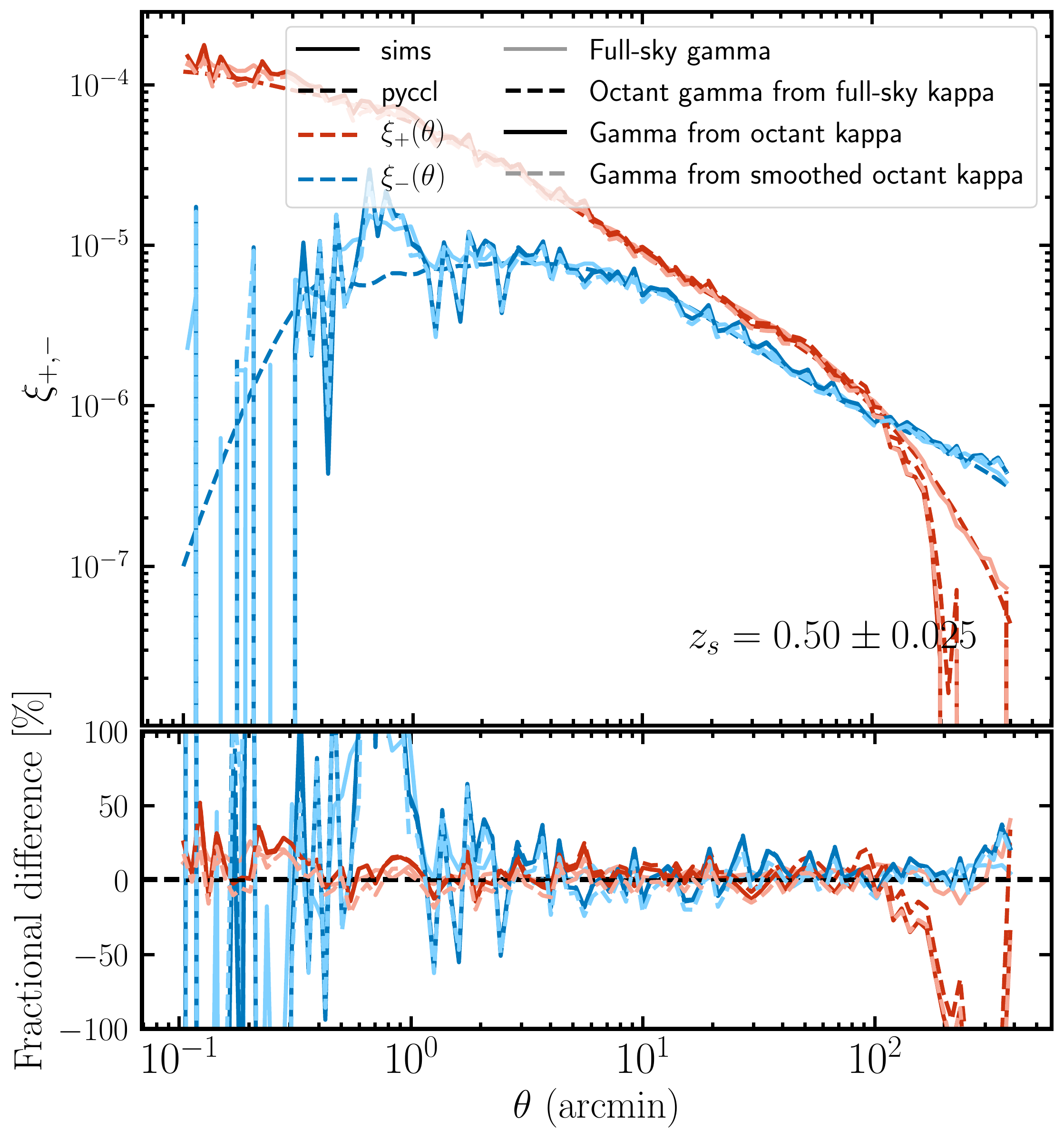}
    \caption{\textbf{Test 1}, described in Appendix~\ref{app:shear}, of the shear-shear auto-correlation, $\xi_\pm (\theta)$, at source redshift of $z_s = 0.5$, using the \texttt{AbacusSummit\_huge\_c000\_ph201}simulation. The source catalogue is obtained using downsampled halo and particle catalogues. The full-sky gamma measurements of $\xi_\pm (\theta)$ (pale solid lines) agree best with the theoretical prediction, as they do not suffer from boundary effects. We only see noticeable differences with the other curves for $\theta > 100$ arcmin. The maps obtained by converting an octant of $\kappa(\hat n)$ yield almost identical results regardless of whether a smoothed mask is applied  or not (dark solid and pale dashed lines). The $\gamma$ measurements cut out from the full-sky $\gamma$ maps (dark dashed lines) exhibit only marginal improvement compared with the other two octant results, suggesting that the main contributor to the noise is sample variance at those scales.}
    \label{fig:xipm_v2}
\end{figure}

\begin{figure}
    \centering
    \includegraphics[width=0.48\textwidth]{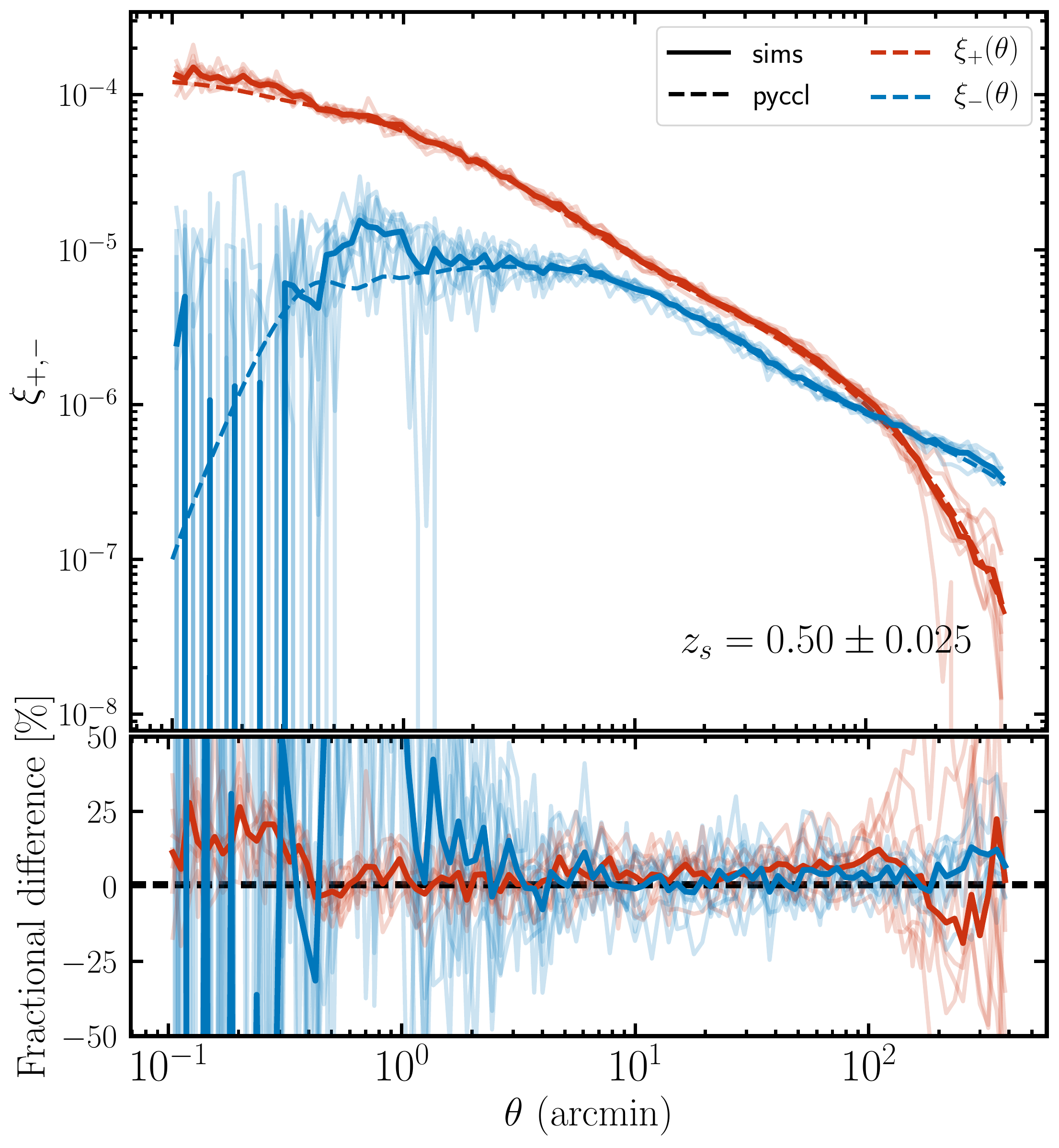}
    \includegraphics[width=0.48\textwidth]{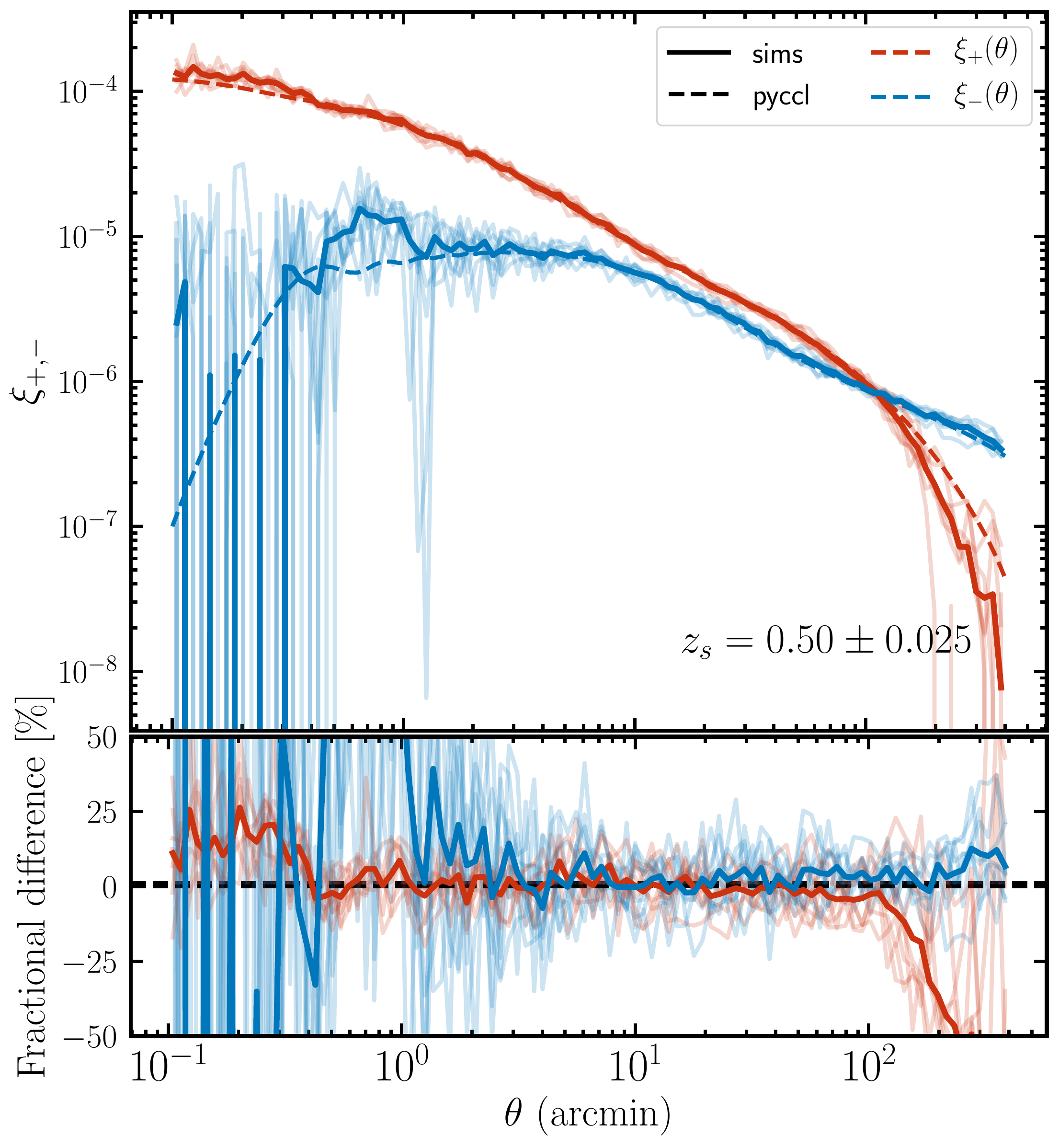}
    \caption{\textbf{Test 2}, described in Appendix~\ref{app:shear}, of the shear-shear auto-correlation, $\xi_\pm (\theta)$, at source redshift of $z_s = 0.5$, using the \texttt{AbacusSummit\_huge\_c000\_ph201}simulation. This test serves as a check of the effect of converting an octant $\kappa (\hat n)$ map into $\gamma_{1,2} (\hat n)$. The top panel shows the idealized scenario in which the $\gamma_{1,2} (\hat n)$ map is obtained from the full-sky $\kappa (\hat n)$ map; then, the full-sky $\gamma_{1,2} (\hat n)$ map is partitioned into the 8 octants. The lower panel corresponds to the case in which the full-sky $\kappa (\hat n)$ map is first partitioned and then each of the eight octants is converted into $\gamma_{1,2} (\hat n)$. The $\xi_-$ curves appear to be largely unchanged, whereas the $\xi_+$ curves are affected beyond $\theta > 100$ arcmin: the average curve in the lower panel displays a cutoff, much like the one in Fig.~\ref{fig:xipm}, whereas the average curve in the upper one is more consistent with \texttt{pyccl} and the full-sky result from Fig \ref{fig:xipm_v2}. On scales beyond $\theta > 100$, we are dominated by cosmic variance.}
    \label{fig:xipm_v1}
\end{figure}

In Fig.~\ref{fig:xipm_v1}, we see that the $\xi_-$ curves appear to be largely unchanged when comparing set one with set two, whereas the $\xi_+$ curves differ from each other significantly beyond $\theta > 100$ arcmin. In particular, the average curve in the lower panel displays a cutoff, much like the one in Fig.~\ref{fig:xipm}, whereas the average curve in the upper panel is more consistent with \texttt{pyccl} and the full-sky result from Fig \ref{fig:xipm_v2}. We conclude that the octant conversion introduces noise along the boundaries that decorrelates the signal on large scales for $\theta_+$, $\theta > 100$, which are also strongly dominated by cosmic variance. We also confirm this via a visual inspection of the difference in the $\gamma_{1,2} (\hat n)$ maps obtained via both procedures. The source catalogue is constructed using a heavily downsampled halo and particle catalogues, which contributes to the larger overall variance of the signal compared with Fig.~\ref{fig:xipm}, as the number density of sources is drastically reduced and the randomly chosen halo particles introduce further noise into the measurement.

\section{Comparison of \textsc{AbacusSummit} power spectra to \texttt{halofit}}
\label{app:halofit}

In many of the validation plots throughout this paper, we compare the simulation results from \textsc{AbacusSummit} with \texttt{pyccl}, which adopts the non-linear matter power spectrum from \texttt{halofit}. However, it is not clear whether \texttt{halofit} provides a good fit to the nonlinear clustering on small scales. We address this question in this Appendix by comparing the power spectrum computed from the 25 base-resolution and the two \texttt{huge}-resolution boxes with \texttt{halofit}. The inclusion of the \texttt{huge}-resolution simulations allows us to test the resolution effects of these coarser runs, which as we found in Fig.~\ref{fig:clkk}, appear to contribute non-negligibly to the small-scale signal. 

\begin{figure}
    \centering
    \includegraphics[width=0.48\textwidth]{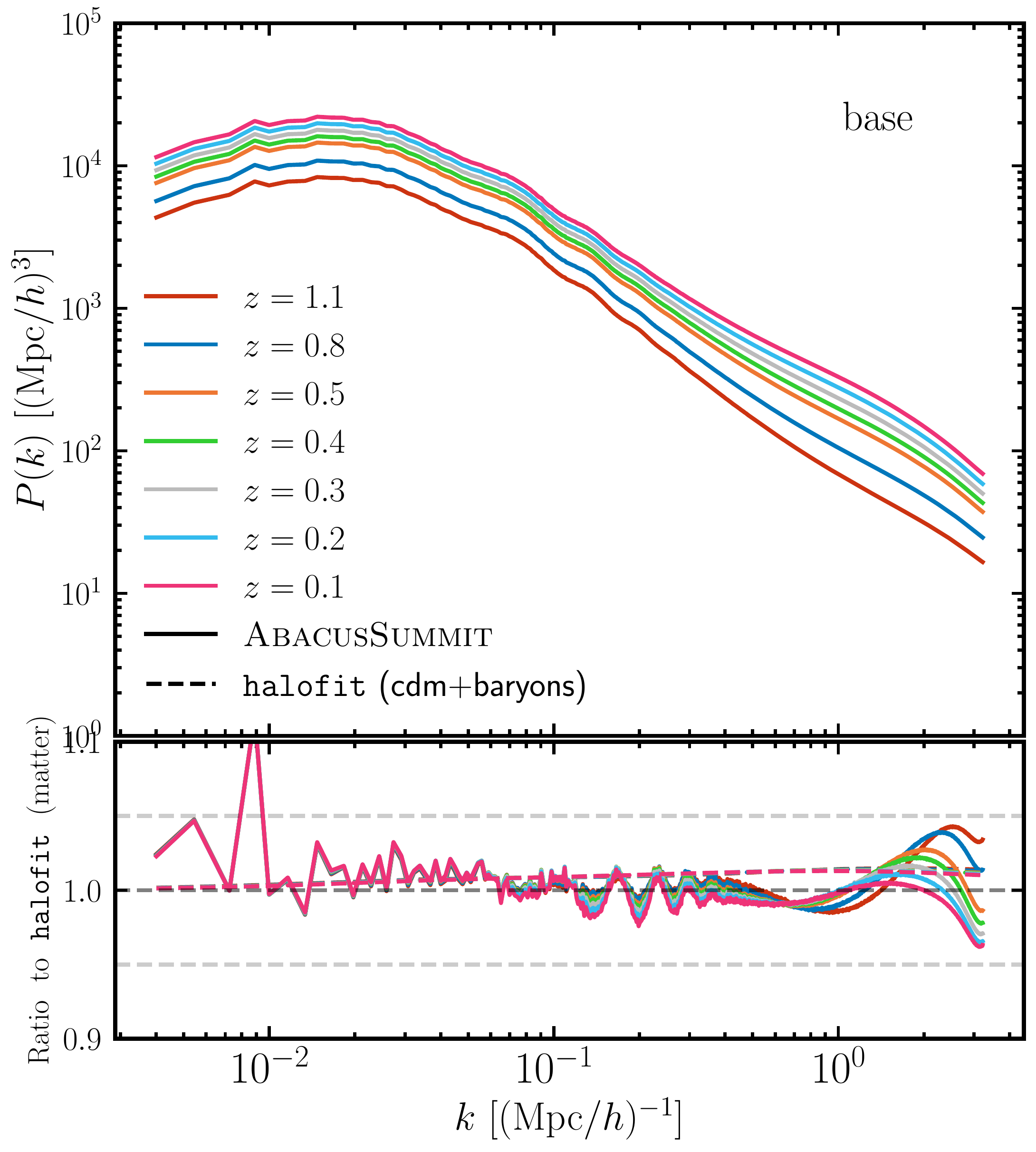}
    \includegraphics[width=0.48\textwidth]{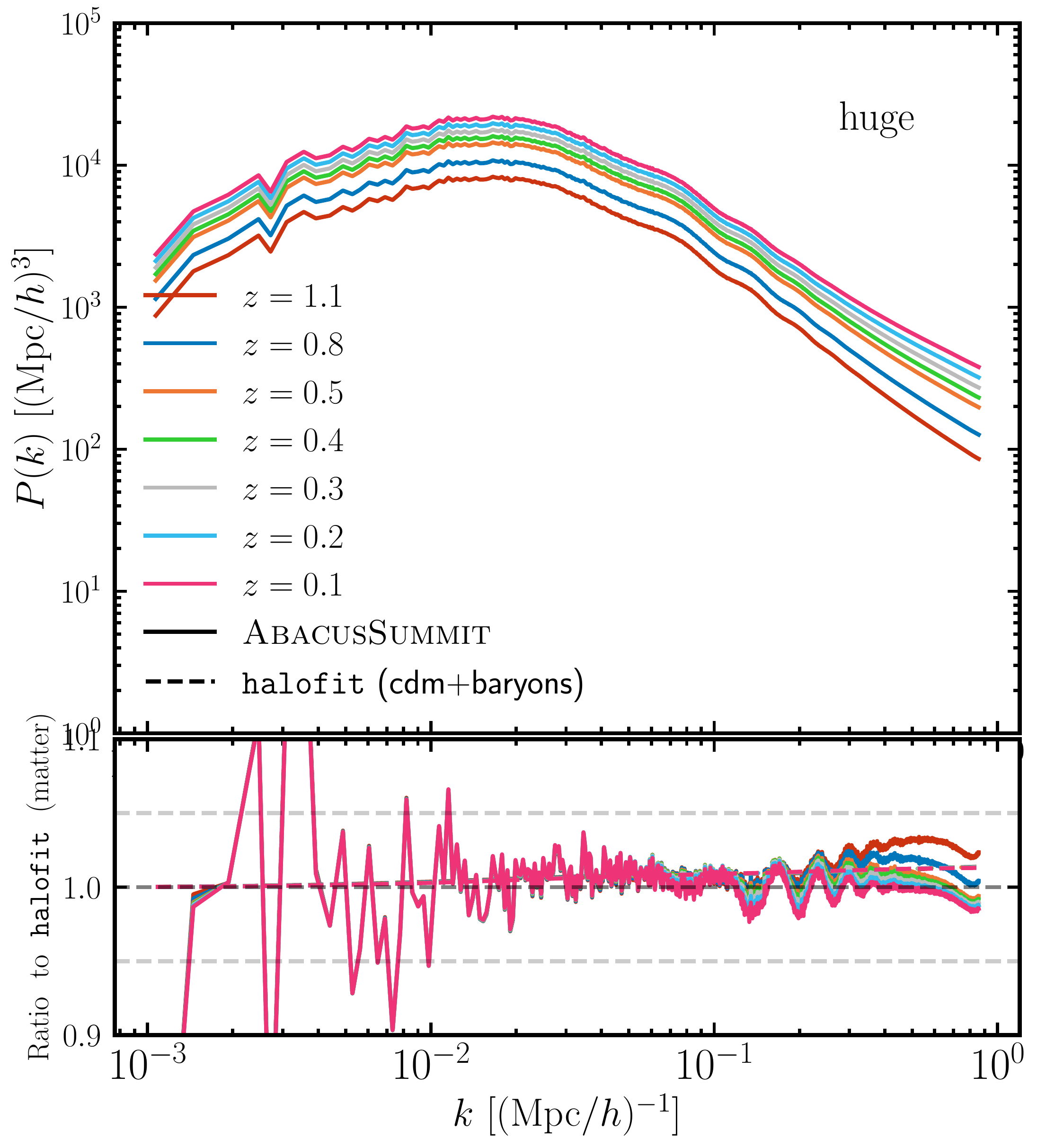}
    \caption{Power spectrum comparison between \textsc{AbacusSummit} and \texttt{halofit} at various redshifts: $z = 0.1$, 0.2, 0.3, 0.4, 0.5, 0.8 and 1.1. The top panel shows the averaged result for the 25 base-resolution boxes, whereas the bottom one comes from the two \texttt{huge}-resolution simulations. The lower segment of each panel shows the ratio of the \textsc{AbacusSummit} power spectrum to the total matter power spectrum obtained via \texttt{halofit}, as well as the ratio of the CDM + baryons \texttt{halofit} power to the total matter \texttt{halofit} power. For all scales shown, the discrepancies are less than 5\%, in agreement with the quoted 5\% error reported for \texttt{halofit} At fixed scale, $k = 1\hompc$, the \texttt{huge} boxes show larger differences with respect to \texttt{halofit} (matter), which explains the lower panel findings in Fig.~\ref{fig:clkk}.}
    \label{fig:halofit}
\end{figure}

Fig. \ref{fig:halofit} shows the power spectrum of \textsc{AbacusSummit} and \texttt{halofit} (both total matter and the CDM + baryons) at several different redshifts: $z = 0.1$, 0.2, 0.3, 0.4, 0.5, 0.8 and 1.1. We expect the particles in the \textsc{AbacusSummit} simulations to be tracers of the CDM + baryons component and therefore, the power spectrum to match the CDM + baryons \texttt{halofit} power spectrum better. In reality, it is hard to say whether that statement holds or not, as the difference between the total matter and the CDM + baryons \texttt{halofit} power spectrum is very small (about 2\% for our choice of 0.06 eV neutrino mass). On large scales, however, there are indications that this is likely the case. It appears that the \textsc{halofit} (CDM + baryons) agreement with \textsc{AbacusSummit} worsens, as we go to lower redshifts, where the \textsc{AbacusSummit} power spectrum becomes more akin to the total matter signal. This makes sense, as the initial conditions are tuned to reproduce the response of matter in the presence of a smooth neutrino component at $z = 1$, as the redshift of interest for current spectroscopic surveys. For all scales considered, the discrepancies are less than 5\% (except for on very large scales, which are dominated by cosmic variance). This is in agreement with the quoted 5\% precision error reported for \texttt{halofit} \citep{2012ApJ...761..152T}. We comment that at fixed scale, $k = 1\hompc$, the \texttt{huge} boxes show larger differences with respect to \texttt{halofit} (matter), which explains the lower panel findings in Fig.~\ref{fig:clkk}. The largest $k$ wavemodes shown correspond to the Nyquist frequency for the 2048 grid cells per dimension used in the power spectrum calculation. 

\bsp	
\label{lastpage}
\end{document}